\documentclass[3p]{elsarticle}
\usepackage{lineno}
\modulolinenumbers[5]

\usepackage[toc,page]{appendix}

\usepackage{amsmath, amsthm, amssymb}
\newcommand{\set}[1]{{Set-#1}}
\newcommand{\ncm}{{n$_\mathrm{eq}$/cm$^2$}}

\begin{document}

\begin{frontmatter}

\title{Radiation Damage Study of SensL J-Series Silicon Photomultipliers Using 101.4~MeV Protons}

\author[ucd]{Alexei Ulyanov \corref{cor1}}
\cortext[cor1]{Corresponding author. E-mail address: alexey.uliyanov@ucd.ie}

\author[ucd]{David Murphy}
\author[ucd]{Joseph Mangan}
\author[esa]{Viyas Gupta}
\author[psi]{Wojciech Hajdas}
\author[ucd2]{Daithi de Faoite}
\author[esa2]{Brian Shortt}
\author[ucd]{Lorraine Hanlon}
\author[ucd]{Sheila McBreen}

\address[ucd]{School of Physics, University College Dublin, Belfield, Dublin 4, Ireland}
\address[esa]{Redu Space Services for European Space Agency, ESEC-Galaxia, 2, Devant les H\^etres, B-6890 Transinne, Belgium}
\address[psi]{Paul Scherrer Institute, Forschungsstrasse 111, CH-5232 Villigen PSI, Switzerland}
\address[ucd2]{School of Mechanical and Materials Engineering, University College Dublin, Belfield, Dublin 4, Ireland}
\address[esa2]{European Space Agency, ESTEC, 2200 AG Noordwijk, The Netherlands}

\begin{abstract}
Radiation damage of J-series silicon photomultipliers (SiPMs) has been studied in the context of using these photodetectors in future space-borne scintillation detectors. Several SiPM samples were exposed to 101.4~MeV protons, with 1~MeV neutron equivalent fluence ranging from $1.27\times10^{8}$~\ncm{} to $1.23\times10^{10}$~\ncm{}. After the irradiation, the SiPMs experienced a large increase in the dark current and noise, which may pose problems for long-running space missions in terms of power consumption, thermal control and detection of low-energy events. Measurements performed with a CeBr3 scintillator crystal showed that after exposure to $1.23\times10^{10}$~\ncm{} and following room-temperature annealing, the dark noise of a single 6~mm square SiPM at room temperature increased from 0.1~keV to 2~keV. Because of the large SiPM noise, the gamma-ray detection threshold increased to approximately 20~keV for a CeBr3 detector using a 4-SiPM array and 40~keV for a detector using a 16-SiPM array. Only a small effect of the proton irradiation on the average detector signal was observed, suggesting no or little change to the SiPM gain and photon detection efficiency.

\end{abstract}

\begin{keyword}
\texttt{ silicon photomultiplier \sep SiPM \sep radiation hardness \sep scintillation detector \sep gamma-ray astronomy \sep space \sep EIRSAT-1}
\end{keyword}

\end{frontmatter}


\section{Introduction}

Silicon photomultipliers (SiPMs) are high-gain semiconductor photodetectors that are now replacing traditional photomultiplier tubes (PMTs) in the majority of terrestrial applications, thanks to their compactness, robustness, insensitivity to magnetic fields and low operating voltage. They have also been proposed for many recent space missions and are being used in a gain control system on board the Hard X-ray Modulation Telescope (HXMT), a Chinese X-ray space observatory launched in June 2017~\cite{li2017_hxmt}.

One of the future gamma-ray astronomy missions that will employ and test SiPMs in space is the Educational Irish Research Satellite 1 (EIRSAT-1)~\cite{murphy2018b,murphy2018c}. It is a 2U CubeSat being developed by students at University College Dublin (UCD) as part of the European Space Agency's Fly Your Satellite! (FYS) programme. The satellite will be deployed from the International Space Station (ISS) and is expected to stay in orbit  (405~km altitude, 51.6$^\circ$ inclination) for about one year. One of the satellite's payloads will be a small gamma-ray detector, called the Gamma-ray Module or GMOD. It is designed for detection of gamma-ray bursts and covers an energy range of 50~keV to 1~MeV. The design of the detector is based on earlier studies at UCD using SiPMs for readout of scintillator crystals~\cite{ulyanov2016,ulyanov2017}. The GMOD detector consists of a cuboid 25~mm by 25~mm by 40~mm CeBr3 scintillator crystal connected to a custom-built $4\times4$ SiPM array. The array is composed of 16 MicroFJ-60035-TSV SiPMs, the 6~mm J-series devices from SensL, Ireland (now part of ON Semiconductor). Because of its high hygroscopicity, the CeBr3 crystal is hermetically encapsulated in an aluminium housing with a quartz window on one side for output of scintillation light.

Properties of CeBr3 scintillator crystals are extensively discussed in Ref.~\cite{quarati_cebr3}. CeBr3 is a very bright scintillator material, with a typical light yield of encapsulated crystals being about 40-45 ph/keV. The scintillation emission is characterised by a short decay time of 17~ns. Combined with optimal photodetectors, the scintillator delivers gamma-ray energy resolution of about 4\% at 662~keV. It is also known to be radiation resistant, exhibiting no degradation in energy resolution or light yield after irradiation with 61--184 MeV protons to a fluence level of 10$^{12}$ protons/cm$^2$~\cite{quarati_cebr3}.

In contrast, SiPMs are known to receive damage from proton radiation and their radiation tolerance is of major concern for space applications.  SiPM radiation damage has been the subject of many recent studies~\cite{bohn2009,angelone2010,qiang2013,garutti2014,musienko2015,heering2016,xu2014,li2016,lacombe2018}, particularly in the context of future high energy physics experiments. However, none of the studies provide sufficient information on the radiation tolerance of the J-series SiPMs, the latest generation of SensL SiPMs which will be used in the GMOD detector and are being considered for other gamma-ray astronomy missions (e.g. AMEGO~\cite{woolf2019} and MERGR~\cite{mergr}). Moreover, the majority of the papers are focused exclusively on the SiPM properties and it is often difficult to conclude to what extent the performance of a scintillator based gamma-ray detector would be affected by the SiPM radiation damage. The purpose of this study is to assess potential damage to the J-series SiPMs caused by proton radiation in space and a consequent change in the performance of gamma-ray spectroscopy for EIRSAT-1 and other gamma-ray missions.

\section{Radiation damage effects in SiPMs}
A single SiPM is composed of an array of small avalanche photodiodes (APDs), typically 10 to 50 $\mu$m in size, connected in parallel and known as SiPM microcells or pixels. The APDs operate in Geiger mode: a free charge carrier generated in or drifted to the high-field depletion region (amplification region) of the APD is accelerated by the electric field and creates via impact ionisation a diverging avalanche of electrons and holes. The avalanche is quenched when the voltage across the APD is reduced to the breakdown voltage $V_\mathrm{br}$, the minimum voltage required to maintain an ongoing avalanche. This is typically achieved using a quenching resistor connected in series with each APD. The APD then gradually recharges to the nominal operating voltage $V_\mathrm{op}$ drawing a fixed charge,  which is approximately proportional to the SiPM overvoltage $V_\mathrm{ov} = V_\mathrm{op} - V_\mathrm{br}$ and independent of the energy and number of particles triggering the avalanche. The total SiPM signal is given by the number of fired microcells (multiplied by the signal of a single microcell) and thereby related to the number of incident photons. Because of the finite number of available microcells, the SiPM response to a light pulse is non-linear and may be saturated. The probability for a single incident photon to fire at least one microcell is called the photon detection efficiency (PDE) of the SiPM and is given by the product of the quantum efficiency,  the microcell fill factor (fraction of the SiPM surface filled with APDs) and the probability for a generated photoelectron to start an electron-hole avalanche.

\begin{figure}[h]
\begin{center}
  \includegraphics[width=0.9\textwidth]{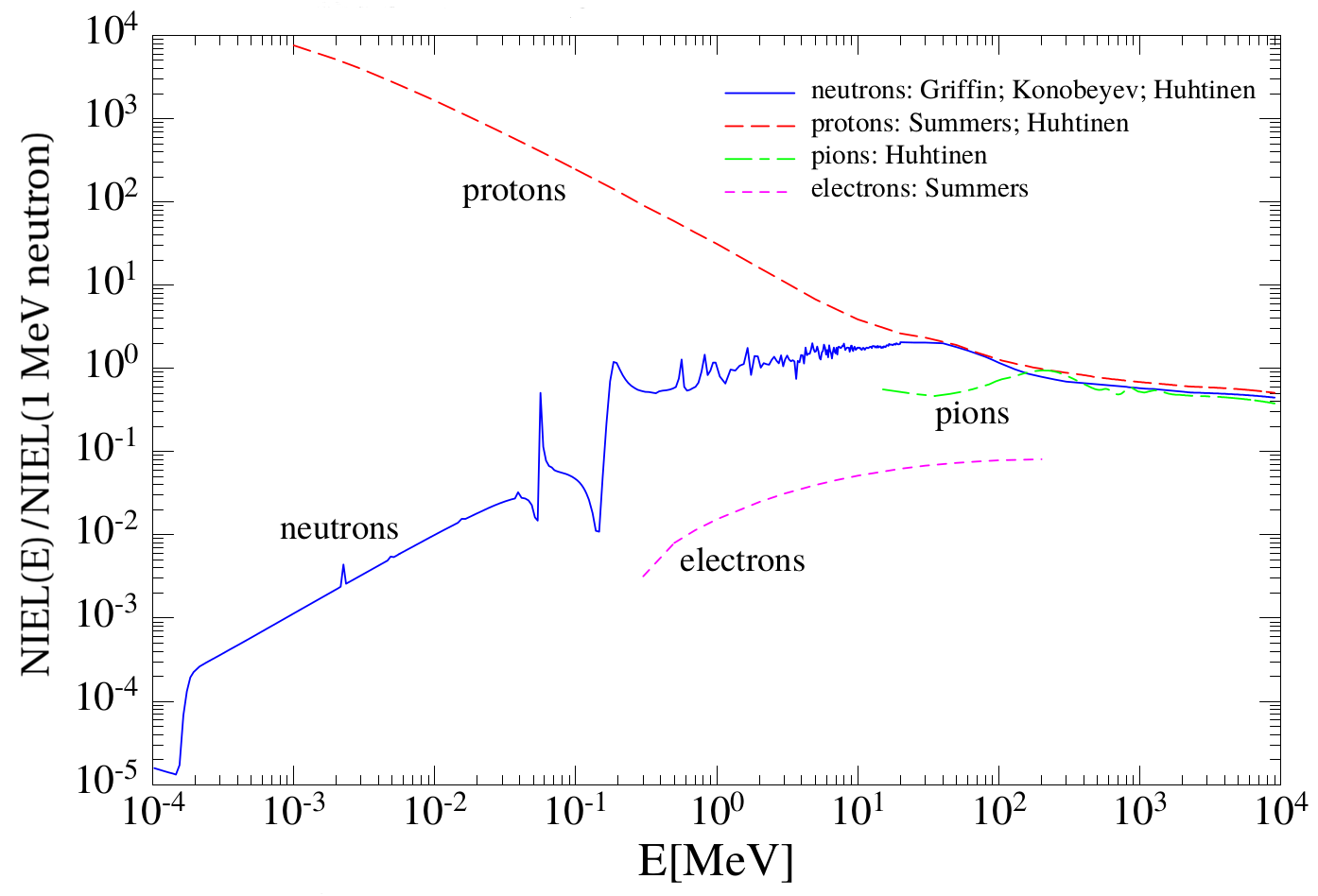}
  \caption{Non-ionising energy loss in silicon for different particles relative to 1~MeV neutrons. Data summarised by Lindstroem~\cite{lindstroem2003} based on~\cite{summers1993,huhtinen1993,griffin1993,konobeyev1992}.}
  \label{fig:displacement_damage}
\end{center}
\end{figure}

As further discussed in Section~\ref{sec:environment}, detector exposure in typical low Earth orbits can reach multiples of 10$^{10}$ protons/cm$^2$ over 10 years of mission operation. An extensive review of various radiation effects in SiPMs is given in Ref.~\cite{garutti2019}. The dominant effect observed in all studies is a significant increase in the SiPM dark count rate and dark current due to displacement of silicon atoms in the crystal lattice caused by hadrons or high-energy electrons ($> 260$~keV). The crystal defects create additional energy levels in the band gap facilitating thermal generation of electron-hole pairs, which generate SiPM dark counts.  The number of crystal defects produced in silicon by energetic particles is typically assumed to be proportional to the non-ionising energy loss (NIEL) of the particles~\cite{lindstroem2003}, although deviations from this approximation have been observed~\cite{huhtinen2002}. For evaluation of displacement damage, it is common to normalise the NIEL of different particles to that of 1~MeV neutrons (Figure~\ref{fig:displacement_damage}) and to express the particle fluence as an equivalent 1~MeV neutron fluence.

Dark counts form background for measurements of light pulses by the SiPM and their statistical fluctuations appear as noise. The dark count rate and noise increase after radiation exposure, which can degrade the SiPM pulse-height resolution and prevent detection of small light pulses.

Fired SiPM microcells need time to recharge before they can efficiently operate. A high dark count rate increases the number of microcells that have been fired and not fully recovered. When such microcells are fired again by an incident light pulse, they produce a signal with a reduced gain. Therefore the SiPM response to light can be reduced or even saturated after radiation exposure if the dark count rate becomes very high.

High radiation exposure can also affect the effective doping density in silicon and change the SiPM breakdown voltage. For example, a 4~V shift in the breakdown voltage has been reported in Ref.~\cite{heering2016} for 10~$\mu$m microcell Hamamatsu SiPMs exposed to proton radiation with 1~MeV neutron equivalent fluence of  $2\times10^{14}$~\ncm.

Apart from the displacement damage in silicon, radiation can also produce ionising damage in the SiO$_2$ layer (or SiO$_2$-Si interface), which covers the surface of the silicon and insulates it from the bias lines and quench resistors. This type of damage results in surface-generated current, which can significantly increase the total SiPM leakage current below the breakdown voltage. Normally, it does not increase the dark count rate and makes little contribution to the dark current above the breakdown voltage. However, if a fraction of the surface-generated charge carriers reach the amplification region, an increase in the dark count rate and an increase in the dark current above the breakdown voltage may also occur. This effect depends on the SiPM design and has been observed in Hamamatsu S10362-11-050C SiPMs irradiated with X-rays to 200~Mrad and 2~Grad~\cite{xu2014}. For proton radiation, however, the effects of displacement damage in SiPMs typically far exceed the effects of ionising damage.

\section{Radiation Environment in Low Earth Orbits}
\label{sec:environment}
Radiation exposure of a satellite in a low Earth orbit (LEO) is dominated by energetic protons in the region of the South Atlantic Anomaly (SAA) and electrons in a high-latitude region where the satellite passes through the outer Van Allen radiation belt. Although the electron fluxes are relatively high, they make little contribution to displacement damage of semiconductor devices (Figure~\ref{fig:displacement_damage}). Therefore only the proton radiation in the SAA region was considered below. The proton exposure of a satellite depends on the orbit altitude and orbit inclination which defines how much time the satellite spends in the SAA region.

In order to estimate potential SiPM exposure to proton radiation in a satellite orbit, the annual proton fluence and average proton spectrum was calculated using ESA's Space Environment Information System (SPENVIS) and the AP-8 model of the trapped proton radiation for solar minimum~\cite{sawyer1976_ap8}. Monte-Carlo simulations using the Geant4 toolkit~\cite{geant4} were then conducted to take into account the shielding effect of the satellite and detector components and to estimate the spectrum and fluence of the protons that pass through the SiPMs.

\begin{figure}
\begin{center}
  \includegraphics[width=0.8\textwidth]{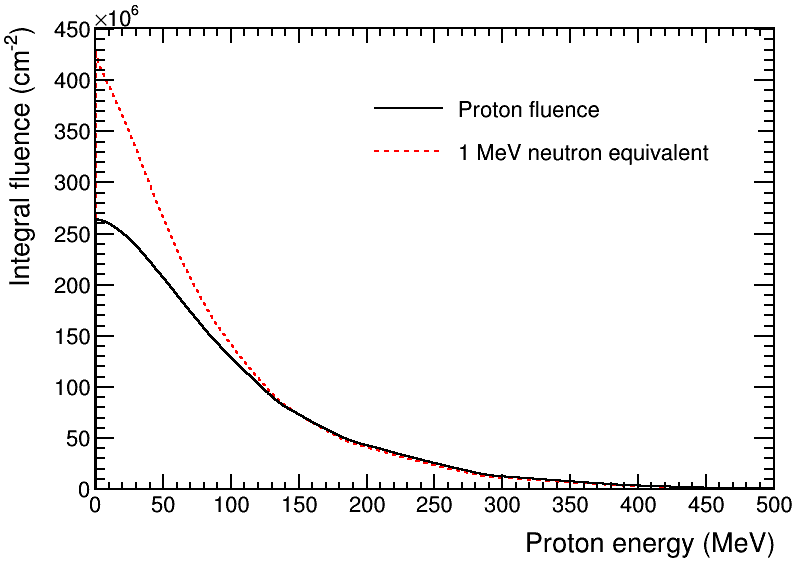}
  \caption{The integral spectrum of protons impinging the SiPM detectors in EIRSAT-1 for one year in the ISS orbit.}
  \label{fig:fluence_iss_diff}
\end{center}
\end{figure}

An integral proton spectrum obtained from the simulations for the ISS orbit is shown in Figure~\ref{fig:fluence_iss_diff}. The average energy of protons that reach the SiPMs is 120 MeV and the total annual omnidirectional fluence is $2.6\times 10^8$~p/cm$^{2}$. The figure also shows 1~MeV neutron equivalent fluence which accounts for the dependence of the non-ionising energy loss on proton energy.

Similar simulations were performed for a circular orbit at 550 km altitude  and for a sun-synchronous orbit (SSO) at 750 km altitude which may be of interest for future missions. The results are summarised in Table~\ref{table:fluence}. For detectors in such orbits, the 1~MeV neutron equivalent fluence can reach  $3\times 10^{10}$~\ncm{} over 10 years of operation (and more, depending on the shielding effect of the satellite).

\begin{table}[h]
  \begin{center}
  \caption{Annual omnidirectional proton fluence expected for the SiPMs in the EIRSAT-1 satellite for different orbits.}
  \label{table:fluence}

  \begin{tabular}{c c c c c}
    \hline
    Orbit & Altitude  & Inclination &  Proton & 1~MeV neutron \\
       &   &    &  fluence & equivalent fluence \\
    \hline
    ISS &   405 km   &  51.6$^\circ$ &  $ 2.6\times 10^8$~p/cm$^{2}$ & $4.3\times 10^8$~\ncm  \\
    LEO &   550 km   &  40.0$^\circ$ &  $ 1.3\times 10^9$~p/cm$^{2}$ & $2.1\times 10^9$~\ncm  \\
    SSO &   750 km   &  98.5$^\circ$ &  $ 2.0\times 10^9$~p/cm$^{2}$ & $3.1\times 10^9$~\ncm  \\

    \hline
  \end{tabular}
  \end{center}
\end{table}

\section{Experimental set-up}

\begin{figure}[h]
\begin{center}
  \includegraphics[width=0.4\textwidth]{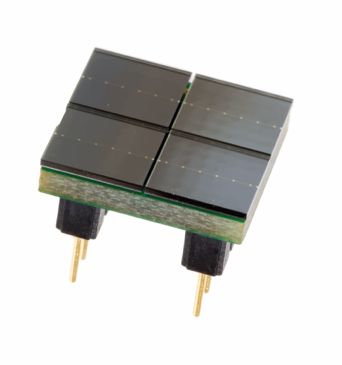}
  \caption{ARRAYJ-60035-4P-PCB board with four MicroFJ-60035-TSV SiPMs (photograph by SensL).}
  \label{fig:sipm}
\end{center}
\end{figure}

Equipment tested in this experiment included two sets of ARRAYJ-60035-4P-PCB SiPM arrays from SensL shown in Figure~\ref{fig:sipm}. Each array comprised four 6~mm MicroFJ-60035-TSV SiPMs. The individual SiPMs were the same model as will be used in the GMOD detector on board EIRSAT-1. Main SiPM parameters are summarised in Table~\ref{table:sipm}.

\begin{table}[h]
  \begin{center}
  \caption{MicroFJ-60035-TSV parameters as specified by the manufacturer~\cite{jseries_ds}. The gain, PDE and crosstalk probability are given at the overvoltage of 3.65~V used in this study. }
  \label{table:sipm}
\begin{tabular}{l l}
  \hline
  Parameter& Value \\
  \hline
  Active area &  6.07$\times$6.07 mm$^2$ \\
  Microcell size & 35 $\mu$m  \\
  No. of microcells &  22292\\
  Breakdown voltage $V_\mathrm{br}$ at 21$^\circ$C & 24.2 to 24.7 V \\
  Temperature dependence of $V_\mathrm{br}$ & 21.5 mV/$^\circ$C\\
  Operating overvoltage $V_\mathrm{ov}$ & 1 to 6 V\\
  Gain &  4$\times$10$^6$ \\
  PDE at 380 nm (peak emission of CeBr3) & 37\% \\
  Capacitance & 4140 pF\\
  Microcell recharge time constant & 50 ns\\
  Crosstalk probability & 13\%\\

  \hline
\end{tabular}
\end{center}
\end{table}

The first SiPM set (\set{1}) comprised four $2\times2$ arrays supplied by SensL in December 2016. Prior to this experiment, they had been used in several studies including an 8-hour high-altitude balloon flight. The second set (\set{2}) included four previously unused $2\times2$ arrays purchased in October 2018.

\subsection{Irradiation of the gamma-ray detector with the \set{1} SiPMs}

The \set{1} SiPMs were used to assemble a gamma-ray detector. The four $2\times2$ arrays were plugged into a passive adapter PCB to form a 16-SiPM array.
The array was then connected to an encapsulated $25\times25\times10$~mm$^3$ CeBr3 scintillator crystal (supplied by Scionix) using standard optical grease from Saint-Gobain Crystals (BC-630) and enclosed with a plastic light-tight housing.

The whole gamma-ray detector including the scintillator and SiPMs was irradiated with 101.4~MeV protons on 6~November 2018. The irradiation was performed at the Proton Irradiation Facility (PIF) at the Paul Scherrer Institute (PSI), Switzerland. PIF is a facility specifically constructed for radiation tests of spacecraft components and includes equipment for beam monitoring and dosimetry. The proton flux during exposures was measured with two ionisation chambers. Both chambers were calibrated using a $2\times 3$~mm$^2$ plastic scintillator mounted at the location of the tested gamma-ray detector prior to the exposures. The same plastic scintillator was used to measure the beam profile.

\begin{figure}
\begin{center}
  \includegraphics[width=0.4\textwidth]{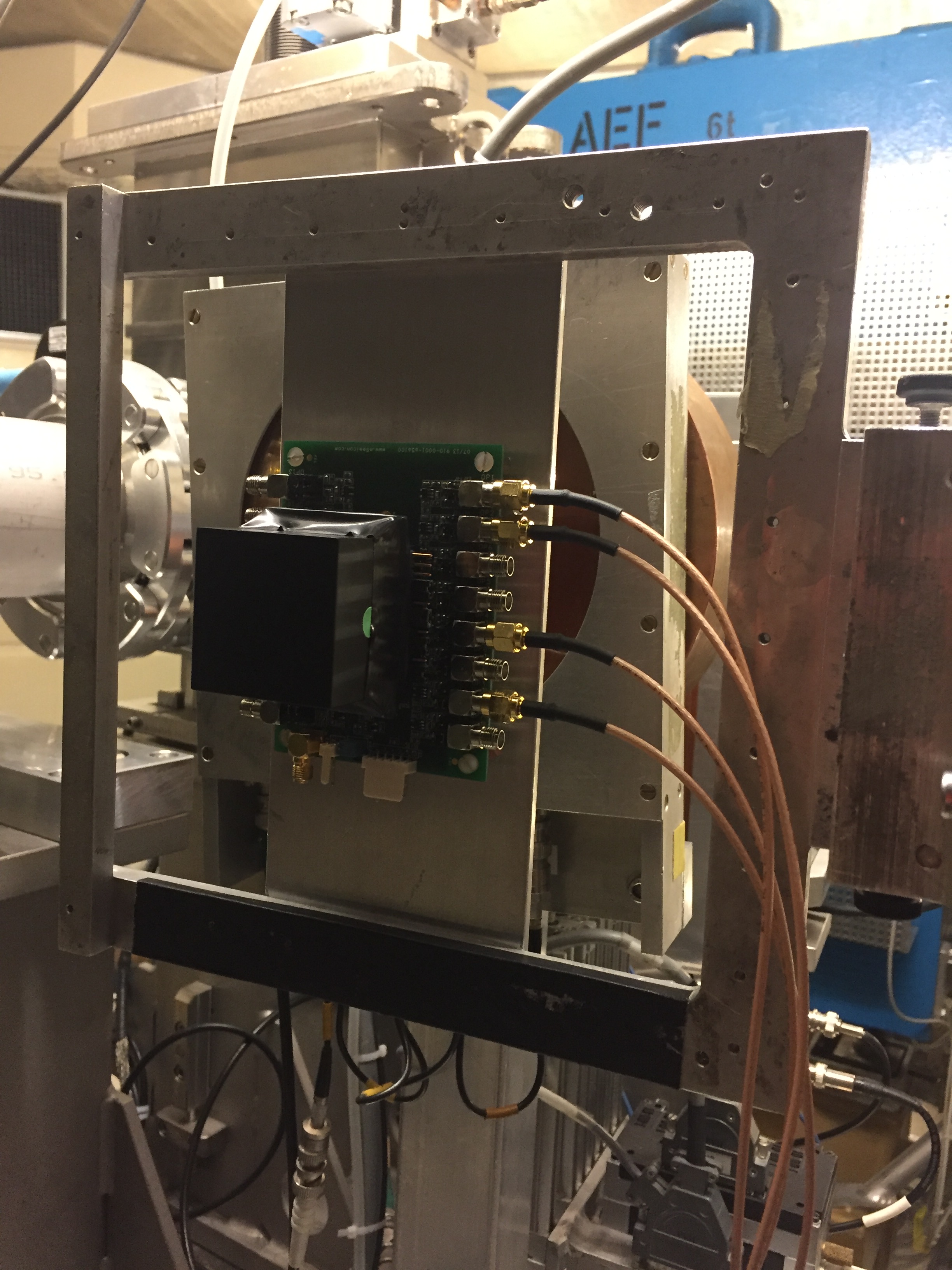}
  \includegraphics[width=0.4\textwidth]{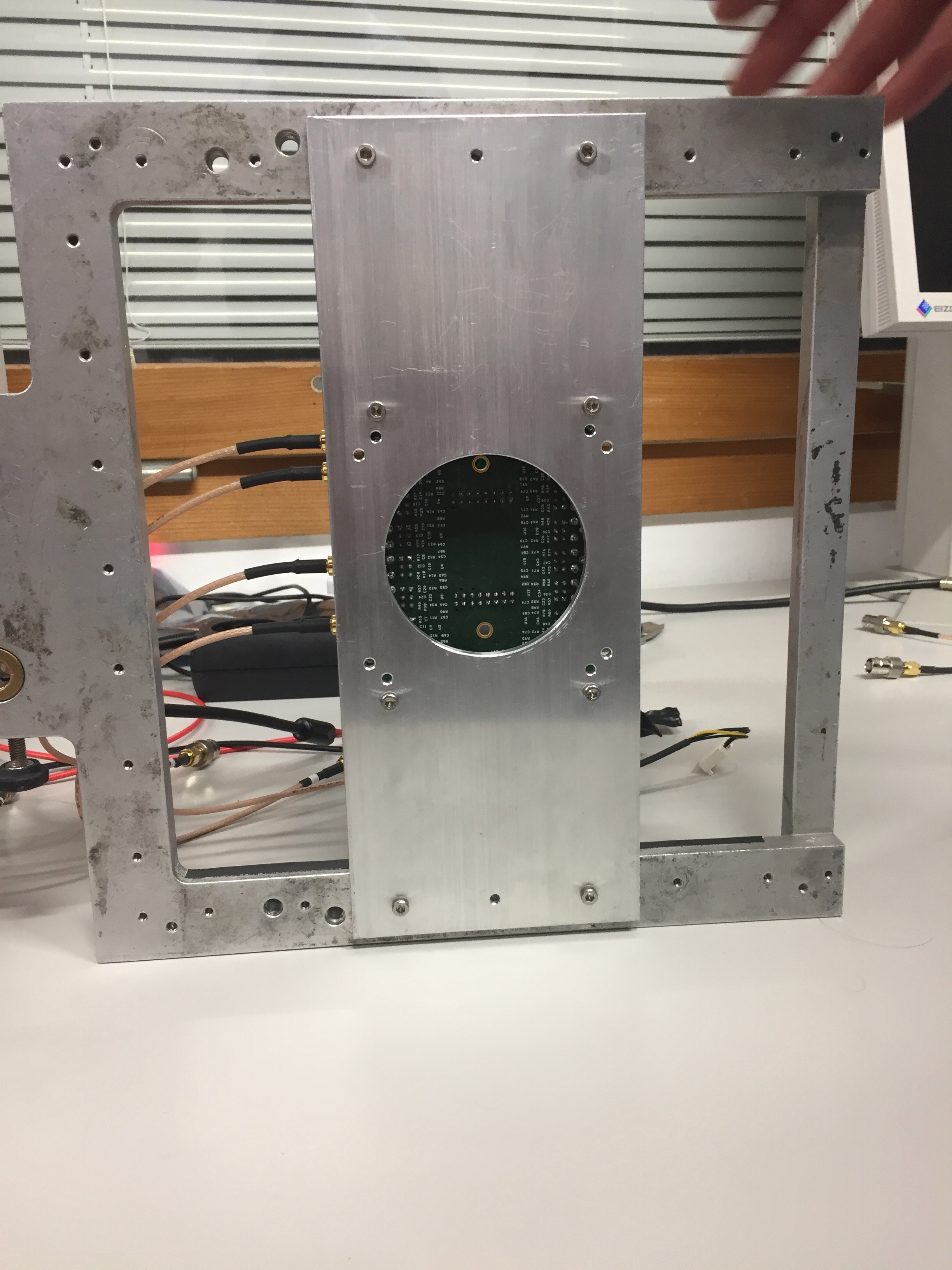}
   \caption{Left: The gamma-ray detector mounted in front of the beam collimator. Right: The back of the detector showing a circular beam entrance window cut out in the aluminium mounting plate.}
  \label{fig:detector_mounted}
\end{center}

\end{figure}

\begin{figure}
  \begin{center}
    \includegraphics[width=0.8\textwidth]{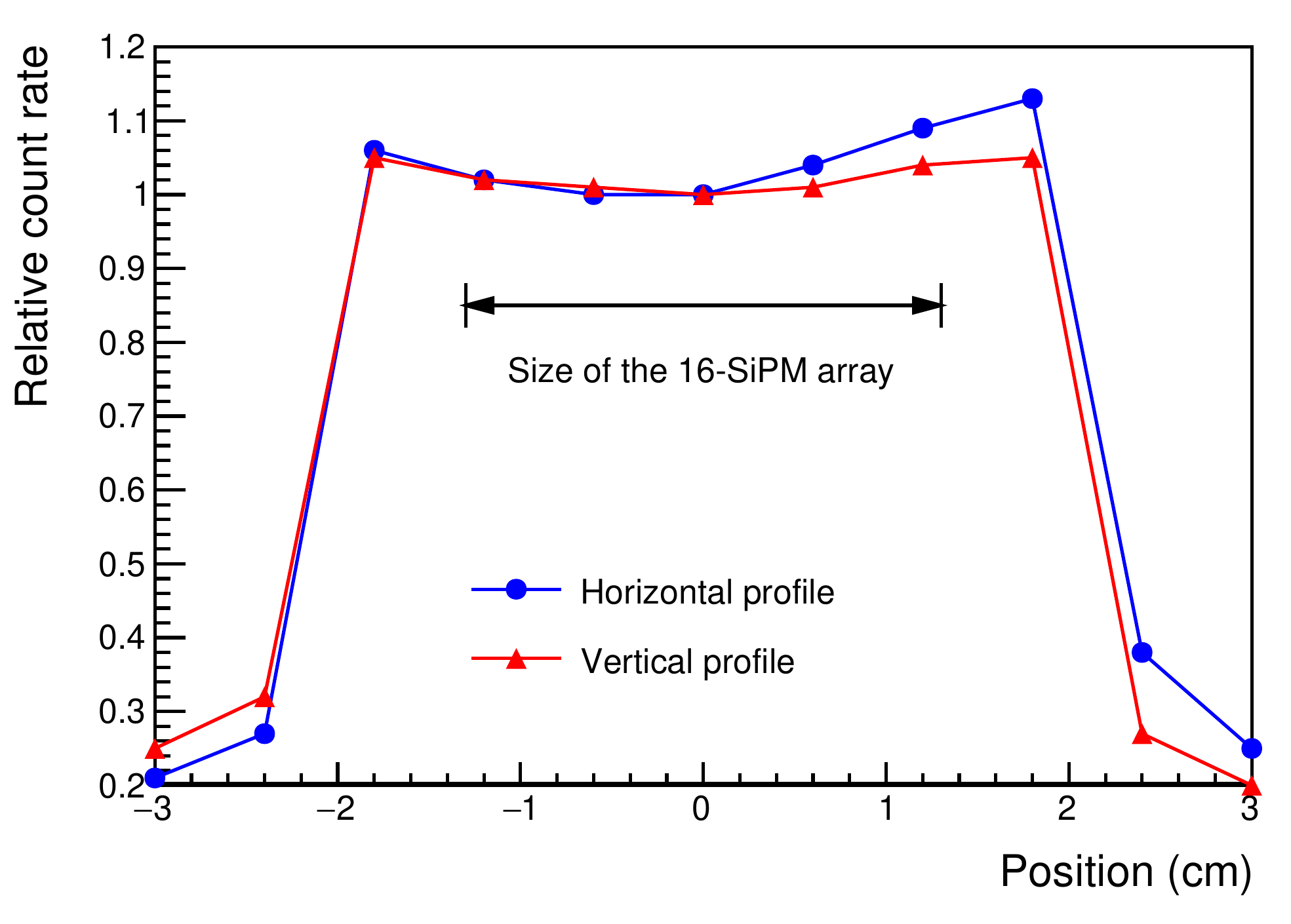}
    \caption{Horizontal and vertical profile of the collimated beam measured with a $2\times 3$~mm$^2$ plastic scintillator counter.}
    \label{fig:beam_profile}
  \end{center}
\end{figure}

The proton beam entered the back of the detector through a circular cut-out in the mounting plate (Figure~\ref{fig:detector_mounted}).
A $40\times40$~mm$^2$ square collimator was used to narrow the beam and limit the exposure and activation of surrounding material. The beam profile at the location of the detector is shown in Figure~\ref{fig:beam_profile}.
The beam was perpendicular to the SiPM plane and crossed four PCBs with a total thickness of 6.4~mm before reaching the SiPMs. Assuming the PCBs were made of FR4 glass-reinforced epoxy with a density of 1.85~g/cm$^3$ and a proton stopping power of 6.5 MeV~cm$^2$/g at 100 MeV (obtained from the NIST Standard Reference Database 124 for a 60:40 mixture of glass and organic material), the proton energy loss in the PCBs was about 7.7 MeV, reducing the beam energy to approximately 94~MeV. This is expected to have a very small effect, increasing the proton non-ionising energy loss in the SiPMs by a few percent.


The detector was irradiated to three increasing fluence levels which are summarised in Table~\ref{table:irradiation1}. A factor of 1.32 was used to convert the 94~MeV proton fluence to 1~MeV neutron equivalent fluence (Figure~\ref{fig:displacement_damage}). The final cumulative fluence is equivalent to the radiation exposure that the SiPMs on board EIRSAT-1 would receive over six years in a 550~km orbit with 40$^\circ$ inclination.

\begin{table}[h]
  \begin{center}
  \caption{Proton irradiation of the gamma-ray detector including \set{1} SiPMs.}
  \label{table:irradiation1}
  \begin{tabular}{c c c c c c}
    \hline
    Exposure &  Flux            & Duration   & Fluence & \multicolumn{2}{c}{Cumulative fluence}\\
    number &  (p/cm$^{2}$/s) & (s)      & (p/cm$^{2}$) & (p/cm$^{2}$) &  (\ncm)\\    \hline
 1 &   $2.56\times10^6$  & 40  & $1.02\times10^8$  &   $1.02\times10^8$  &  $1.35\times10^8$\\
 2 &   $2.54\times10^6$  & 79  & $2.00\times10^8$  &   $3.02\times10^8$  &  $3.99\times10^8$\\
 3 &   $2.66\times10^7$  & 339  & $9.01\times10^9$  &   $9.31\times10^9$ & $1.23\times10^{10}$ \\
        \hline
  \end{tabular}
  \end{center}
\end{table}

The SiPMs were unbiased (powered off) during the irradiation. After each proton exposure the detector was removed from the beam and placed outside the shielded irradiation area, where characterisation measurements were performed.

\subsection{Irradiation of bare SiPMs (\set{2})}

Bare SiPMs from \set{2} were irradiated at PIF on 7~November 2018. One of the $2\times2$ SiPM arrays was not irradiated and served as a reference in further characterisation measurements. Three other arrays were irradiated with 101.4~MeV protons. Each $2\times2$ SiPM array was irradiated separately and to a different fluence level as summarised in Table~\ref{table:irradiation2}.
A factor of 1.26 was used to convert 101.4 MeV proton fluence to 1~MeV neutron equivalent fluence (Figure~\ref{fig:displacement_damage}).

Similar to the \set{1} SiPMs, the \set{2} SiPMs were unbiased during the irradiation.

\begin{table}[h]
  \begin{center}
  \caption{Proton irradiation of \set{2} SiPMs.}
  \label{table:irradiation2}
  \begin{tabular}{c c c c c}
    \hline
    SiPM  &  Flux           & Duration  & \multicolumn{2}{c}{Fluence} \\
    array &  (p/cm$^{2}$/s) &    (s)    & (p/cm$^{2}$) &  (\ncm) \\   \hline
 1 &    $0$                &  0   &  $0$  & $0$  \\
 2 &   $2.89\times10^6$  & 35  & $1.01\times10^8$   &  $1.27\times10^8$\\
 3 &   $2.87\times10^6$  & 105  & $3.01\times10^8$  &  $3.79\times10^8$\\
 4 &   $2.89\times10^6$  & 347  & $1.00\times10^9$  &  $1.26\times10^{9}$ \\
    \hline
  \end{tabular}
  \end{center}
\end{table}

\subsection{SiPM readout with SIPHRA}
\label{sec:siphra}

SIPHRA is a low-power ASIC recently developed by IDEAS (Norway) for SiPM readout in space applications~\cite{siphra,siphra_ulyanov2017}. It will be employed on board EIRSAT-1 and was used to measure SiPM signals in this study.

The ASIC has 16 input channels. Each input channel has a current-mode input stage (CMIS), which allows DC-coupling of SiPMs and downscales the input signal by a programmable attenuation factor (10, 100, 200 or 400). The downscaled signal is fed into a current comparator and a pulse-height spectrometer. The comparator generates a trigger signal when the input current in any channel exceeds a programmable threshold. The spectrometer consists of a current integrator, an RC-CR pulse shaper with programmable shaping time (200 ns, 400 ns, 800 ns or 1600 ns) and a track-and-hold (TH) unit. The TH unit tracks the shaper output waveform, then stops and holds the signal when it receives the HOLD signal. The HOLD signal is generated internally after a programmable delay (HOLD delay) following the trigger signal. The output of the TH unit is multiplexed to a 12-bit successive-approximation register ADC, which digitises the signals from any or all 16 channels, as well as the summing channel.

The summing channel sums all the 16 input signals and scales down the sum by a factor of 16. It is useful for measuring and triggering on the total signal of a SiPM array, e.g. when it is coupled to a single monolithic scintillator. Similar to the normal input channels, the summing channel can be used to generate trigger signals and for pulse-height spectroscopy.

In this work, SIPHRA was connected to the summed outputs (common cathodes) of $2\times2$ SiPM arrays. Thus, only four SIPHRA channels were used to read the full 16-SiPM array of the irradiated gamma-ray detector. In this configuration, the SIPHRA summing channel was not very useful and was not used (except where noted otherwise). Trigger signals were generated by individual channels and the total energy registered by the gamma detector in each event was calculated by adding the digitised signals from individual channels.

The CMIS attenuation factor of 100 and the shaping time of 200~ns were used in all measurements.

For the HOLD delay, a default value optimised by IDEAS was used. The HOLD delay is ideally chosen so that the TH unit holds the peak value of the shaped pulse. However, the value of trigger threshold affects the time when the trigger signal is generated (time-walk effect) and consequently the time of the HOLD signal. Therefore, the height of the shaped pulse can be measured before or past its peak, depending on the relative values of the trigger threshold and the input pulse amplitude. Thus, the measured signal depends to some extent on the value of trigger threshold. This effect is particularly strong for small signals comparable with the trigger threshold.

In order to perform pulse-height measurements, the trigger threshold in the comparator should be set sufficiently high above the SiPM dark current. This is required to prevent triggering from fluctuations of the SiPM dark current (SiPM noise) which can easily saturate the maximum readout rate of SIPHRA (3 to 50 kHz depending on the number of channels to read). When SiPMs get damaged by radiation, the trigger threshold needs to be raised to account both for the increased dark current and for the increased noise. SIPHRA has a special dark current compensation mode that can automatically compensate the average dark current, in which case the trigger threshold only needs to account for the SiPM noise. The dark current compensation mode was not used in the measurements described in this report, as the authors were not initially aware of it.

\subsection{Characterisation set-up}
\label{sec:measurements}
Characterisation measurements of the gamma-ray detector with the \set{1} SiPMs were performed in PSI on 6 November 2018 before the proton irradiation and within 10-30 minutes after each exposure.

Following the proton irradiation and a one-month deactivation period, the \set{2} SiPMs and the detector with the \set{1} SiPMs  were shipped back to UCD, where the detector characterisation measurements were repeated, approximately three months after the irradiation. The gamma-ray detector was then disassembled and characterisation of individual $2\times 2$ arrays was performed for both SiPM sets.

\begin{figure}
\begin{center}
  \includegraphics[width=0.8\textwidth]{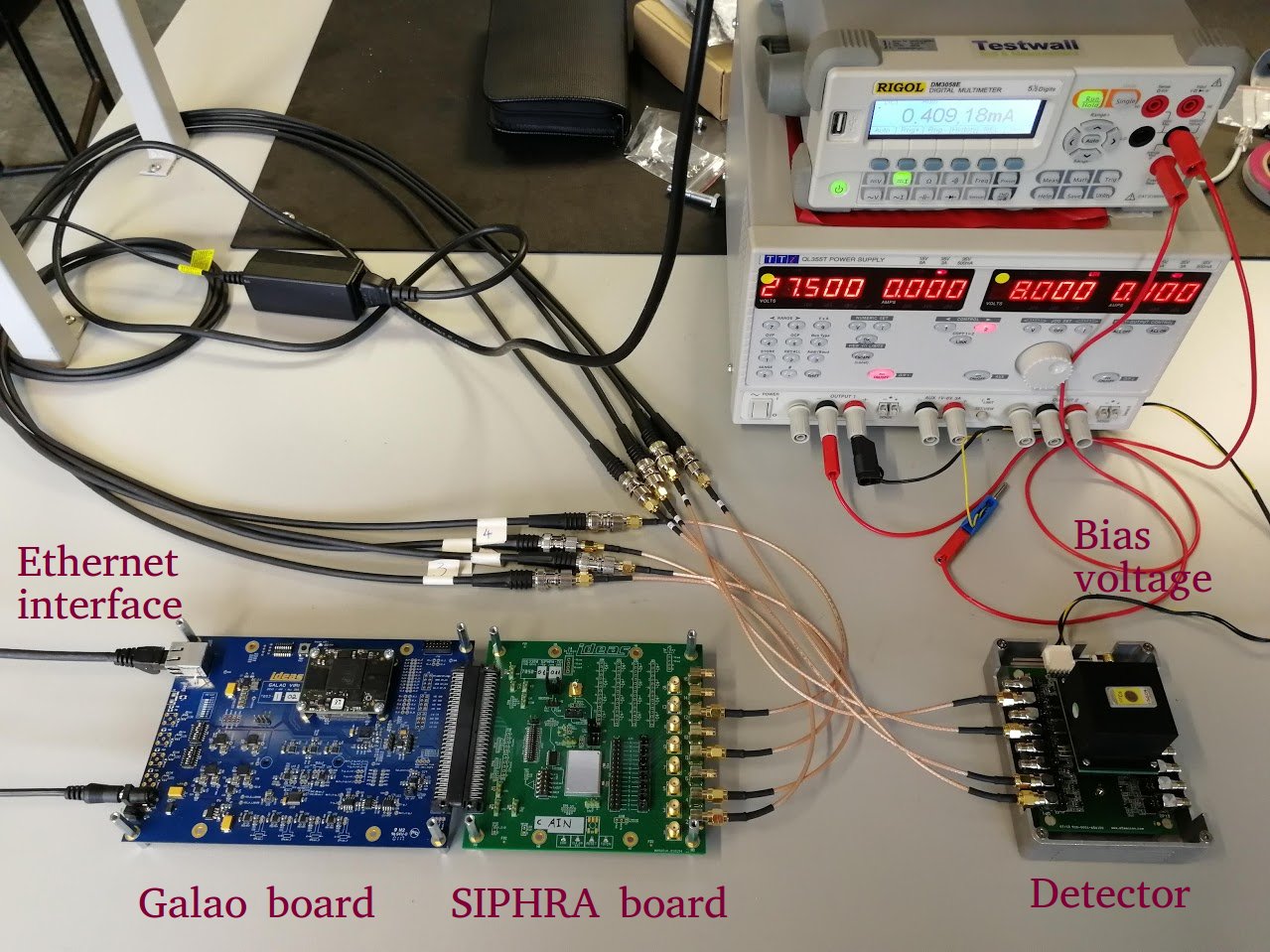}
  \caption{Detector characterisation set-up.}
  \label{fig:measurementsetup}
\end{center}
\end{figure}

The measurement set-up for characterisation of the gamma-ray detector is shown in Figure~\ref{fig:measurementsetup}. The readout board with the SIPHRA ASIC (SIPHRA board supplied by IDEAS) was connected to the detector by four 2.5~m coaxial cables: each of the $2\times 2$ arrays was connected to one SIPHRA input. The SIPHRA board was then connected to a general purpose readout and control board (Galao board supplied by IDEAS) via a board-to-board connector. The Galao board was connected to a PC via the Ethernet interface.
The SiPM bias voltage was supplied by a benchtop power supply unit (PSU) and the SiPM current was measured by a digital multimeter (RIGOL DM3058E in UCD / TENMA 72-7720 in PSI) connected in series with the PSU.

To measure the detector response to gamma rays, a sealed radioactive source containing $^{137}$Cs or $^{241}$Am radionuclide was placed directly on top of the detector housing.

The same set-up was used in UCD to measure the dark current and noise of the \set{2} SiPMs: the four arrays were plugged into the adapter PCB instead of the gamma-ray detector, enclosed with a light-tight housing and connected to the SIPHRA board. To measure the current of each $2\times2$ array, they were connected to the SIPHRA board one at a time.

\begin{figure}[h]
\begin{center}
  \includegraphics[width=0.8\textwidth]{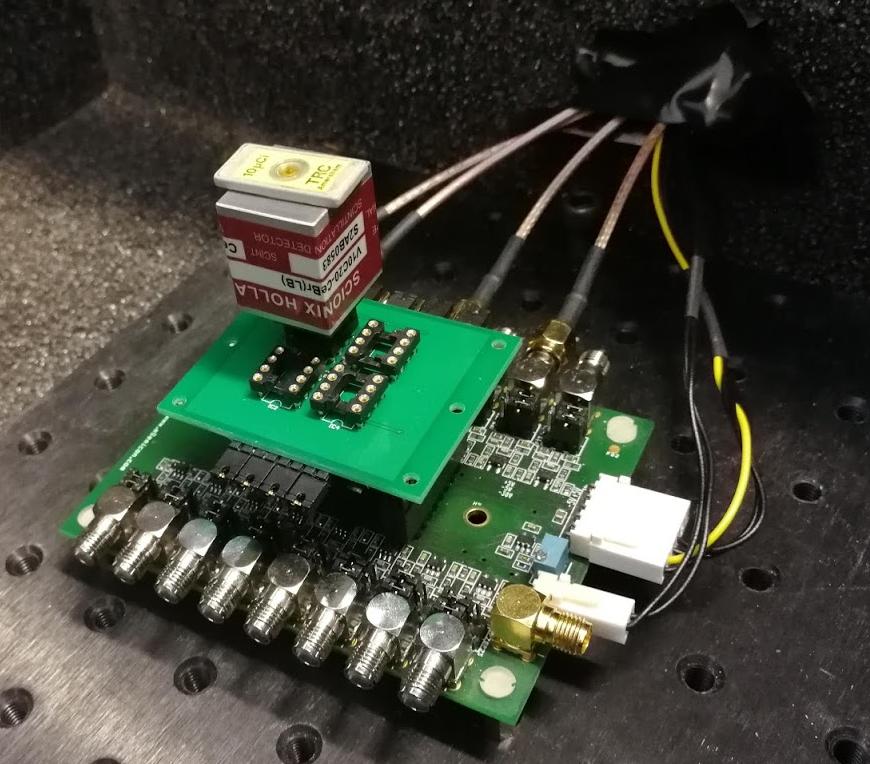}
  \caption{Measurement of the gamma-ray response of the small CeBr3 scintillator coupled to a $2\times 2$ SiPM array.}
  \label{fig:measurement_setup_single}
\end{center}
\end{figure}

In addition, gamma-ray measurements were performed using the $2\times 2$ SiPM arrays individually coupled to a small 10~mm $\times$ 10~mm $\times$ 10~mm CeBr3 scintillator crystal. These measurements were carried out in a light-tight box using the same experimental set-up with just one $2\times 2$ array SiPM array connected to the adapter board (Figure~\ref{fig:measurement_setup_single}).

Except where noted otherwise, all measurements in this study were performed using a SiPM bias voltage of 28.15~V
which corresponds to an overvoltage of 3.65~V.
It should be noted that each $2\times2$ SiPM array was connected to the bias supply through a low-pass filter consisting of a 10~ohm resistor and a 50~nF capacitor. No compensation or correction for the voltage drop across the 10~ohm resistor was used in this work. This effect was only significant for SiPMs irradiated to the highest fluence of $1.23\times10^{10}$~\ncm{}, where depending on temperature the greatly increased SiPM current resulted in a moderately small 50--90~mV voltage drop in the majority of the measurements.

\clearpage
\section{Results}

\subsection{SiPM dark current at $V_\mathrm{bias}=28.15$\,V}
\label{sec:current}

Measurements in PSI were conducted at a temperature of 24$^\circ$C (measured with a conventional liquid-in-glass thermometer). The total dark current of the gamma-ray detector (for all 16 SiPMs) was 100~uA before the proton irradiation. The current rose to 1.50~mA after the first proton exposure of $1.35\times10^8$~\ncm{} (measured within 10 minutes after the irradiation) and then reduced to 1.32 mA in the next 20 minutes indicating a possible recovery of the induced radiation damage. A similar effect was observed after the second exposure: the initial current of 2.6~mA decreased to 2.3~mA in 11 minutes. However, after the third exposure (bringing the cumulative fluence to $1.23\times10^{10}$~\ncm{}), the SiPM current increased to 55~mA upon powering up the SiPMs and continued to grow. This continuous increase of the SiPM dark current after the third exposure was due to SiPM self-heating, as the power dissipated by the SiPMs drawing the current of 55~mA was about 1.5~W.

The detector current measured in UCD 86 days after the irradiation was 20~mA at a temperature of 21$^\circ$C right after switching on the SiPM bias voltage. It then gradually increased to 35~mA in two hours due to the rising temperature of the SiPMs. The temperature measured by a resistance temperature detector (PT100) mounted on the socket adapter board increased from  21$^\circ$C to 36$^\circ$C. The temperature of the SiPMs might have been higher, as the SiPMs were mounted on separate boards plugged into the adapter board. The PT100 sensor was not used in the PSI measurements.

\begin{figure}
  \begin{center}
    \includegraphics[width=0.8\textwidth]{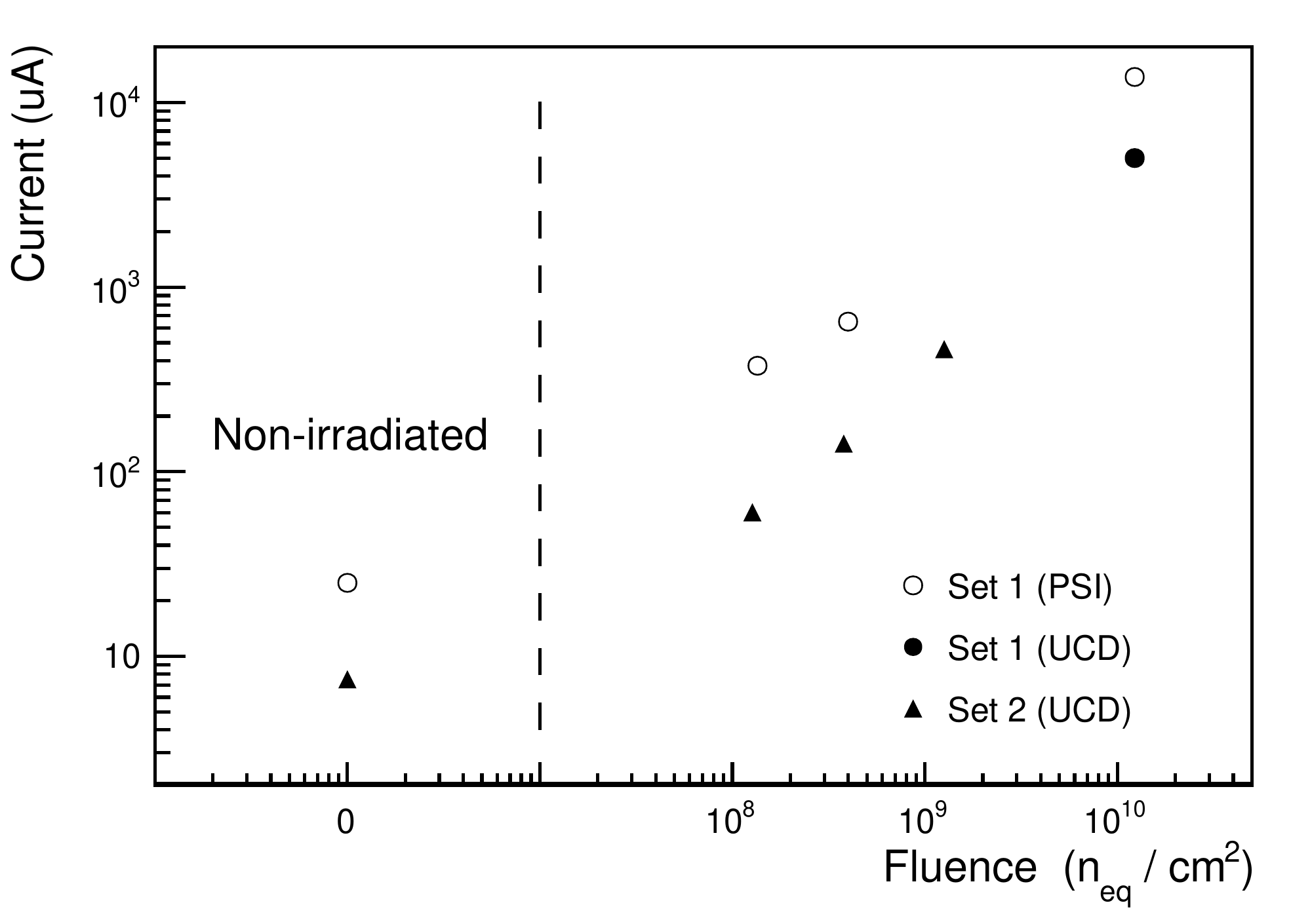}
    \caption{Dark current of a $2\times2$ SiPM array (1.47~cm$^2$ active area) as a function of irradiation fluence. The PSI measurements were performed at 24$^\circ$C about 10 minutes after the SiPM irradiation. The UCD measurements were performed  at 21$^\circ$C 86 days after the SiPM irradiation. The bias voltage was 28.15~V in all measurements.}
    \label{fig:current_vs_fluence}
  \end{center}
\end{figure}

The initial current measured with "cold" SiPMs is shown in Figure~\ref{fig:current_vs_fluence} for both SiPM sets. For \set{1} the current was measured using the 16-SiPM array and divided by 4 to represent the current drawn by one $2\times2$ SiPM array. The current drawn by the \set{2} SiPMs (measured in UCD) turned out to be smaller by a factor of 3 than the current measured for \set{1} after a similar exposure. This can be explained by partial recovery from the radiation damage in the 86 days that passed between the irradiation and the measurements. In addition, it is estimated that the 3$^\circ$C lower temperature in the UCD measurements decreased the dark current by a factor of 1.35 for non-irradiated SiPMs and by a factor of 1.1 for the irradiated SiPMs (see Section~\ref{sec:temperature}). It is not clear why the non-irradiated SiPMs from \set{1} drew much larger current than the non-irradiated SiPMs from \set{2}. This could be related to the age of the sensors or possibly some damage they received in previous studies.

\begin{figure}
  \begin{center}
    \includegraphics[width=0.8\textwidth]{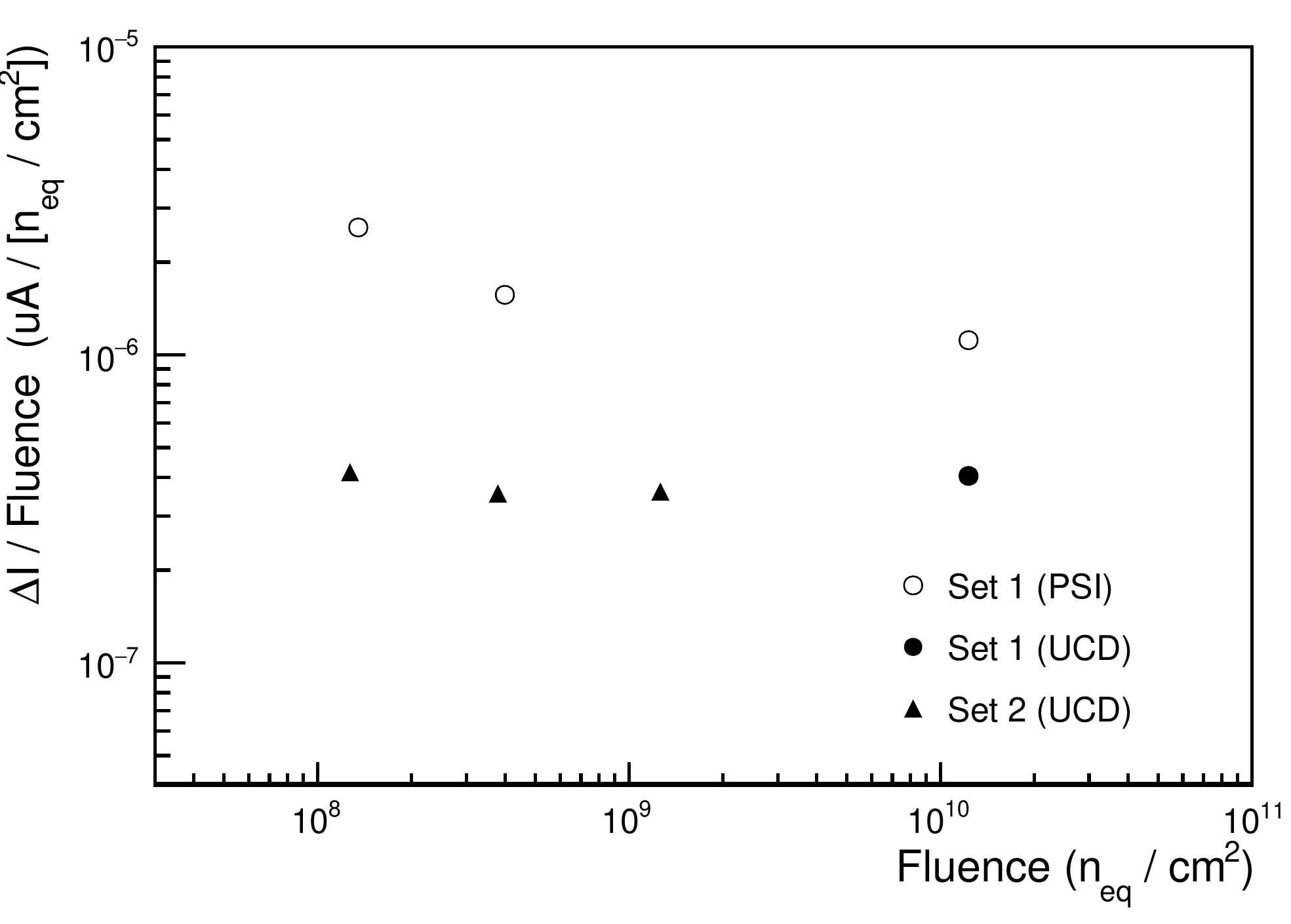}
    \caption{Increase in the dark current of a $2\times2$ SiPM array (1.47~cm$^2$) divided by the cumulative fluence. The PSI measurements were performed  at 24$^\circ$C about 10 minutes after the SiPM irradiation. The UCD measurements were performed  at 21$^\circ$C 86 days after the SiPM irradiation.  The bias voltage was 28.15~V in all measurements.}
    \label{fig:current_over_fluence}
  \end{center}
\end{figure}

The increase in SiPM dark current after irradiation $\Delta I = I_\mathrm{after} - I_\mathrm{before}$ was expected to be proportional to the amount of displacement damage and therefore to the irradiation fluence. As can be seen from Figure~\ref{fig:current_over_fluence}, the results obtained are in reasonable agreement with this hypothesis with the exception of the lowest fluence of $1.35\times10^8$~\ncm{}, where the current drawn by \set{1} SiPMs was surprisingly high. It should be noted that, in the UCD measurements performed 86 days after the irradiation, the \set{1} and \set{2} results were reasonably consistent with each other.

\subsection{SiPM dark current as a function of bias voltage}
\label{sec:iv}

Current-voltage curves of the irradiated $2\times2$ SiPM arrays were measured in UCD on 1 February 2019. No such characterisation was performed prior to irradiation.

\begin{figure}
  \begin{center}
    \includegraphics[width=0.8\textwidth]{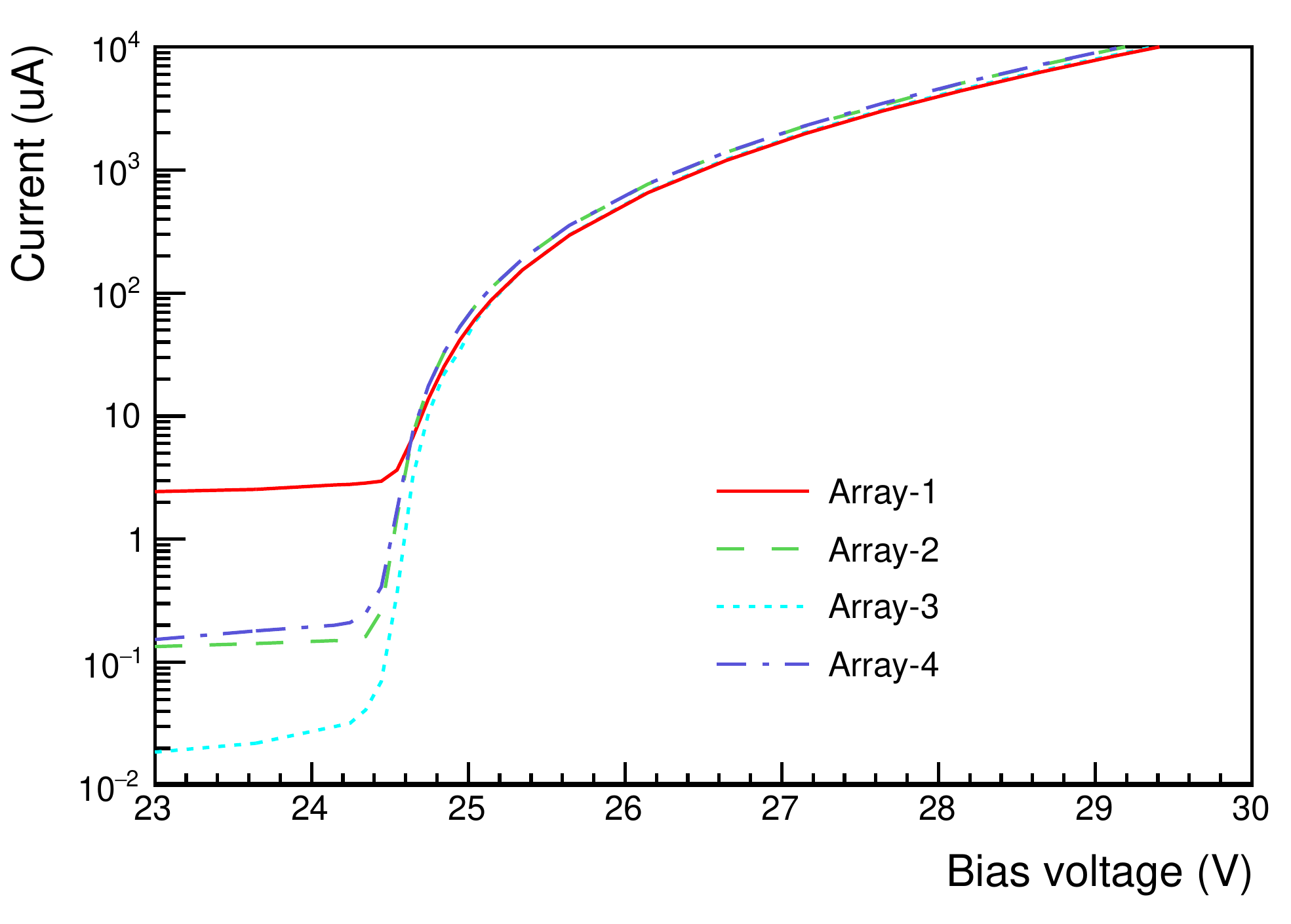}
    \caption{Current-voltage characteristics of the $2\times2$ SiPM arrays (1.47~cm$^2$) from \set{1} after proton irradiation with a cumulative fluence of $1.23\times10^{10}$~\ncm{}. Measured 87 days after the irradiation. $T=21^\circ$C. }
    \label{fig:iv_set1}
  \end{center}
\end{figure}

The curves for the SiPM arrays from \set{1} are shown in Figure~\ref{fig:iv_set1}. All SiPM arrays displayed a sharp rise in current when the bias voltage was increased above the SiPM breakdown voltage ($\sim 24.5$~V). The current of all four arrays was very similar above the breakdown voltage. However, a very large difference was observed below the breakdown voltage, with the current of Array-1 being particularly large.

\begin{figure}
  \begin{center}
    \includegraphics[width=0.8\textwidth]{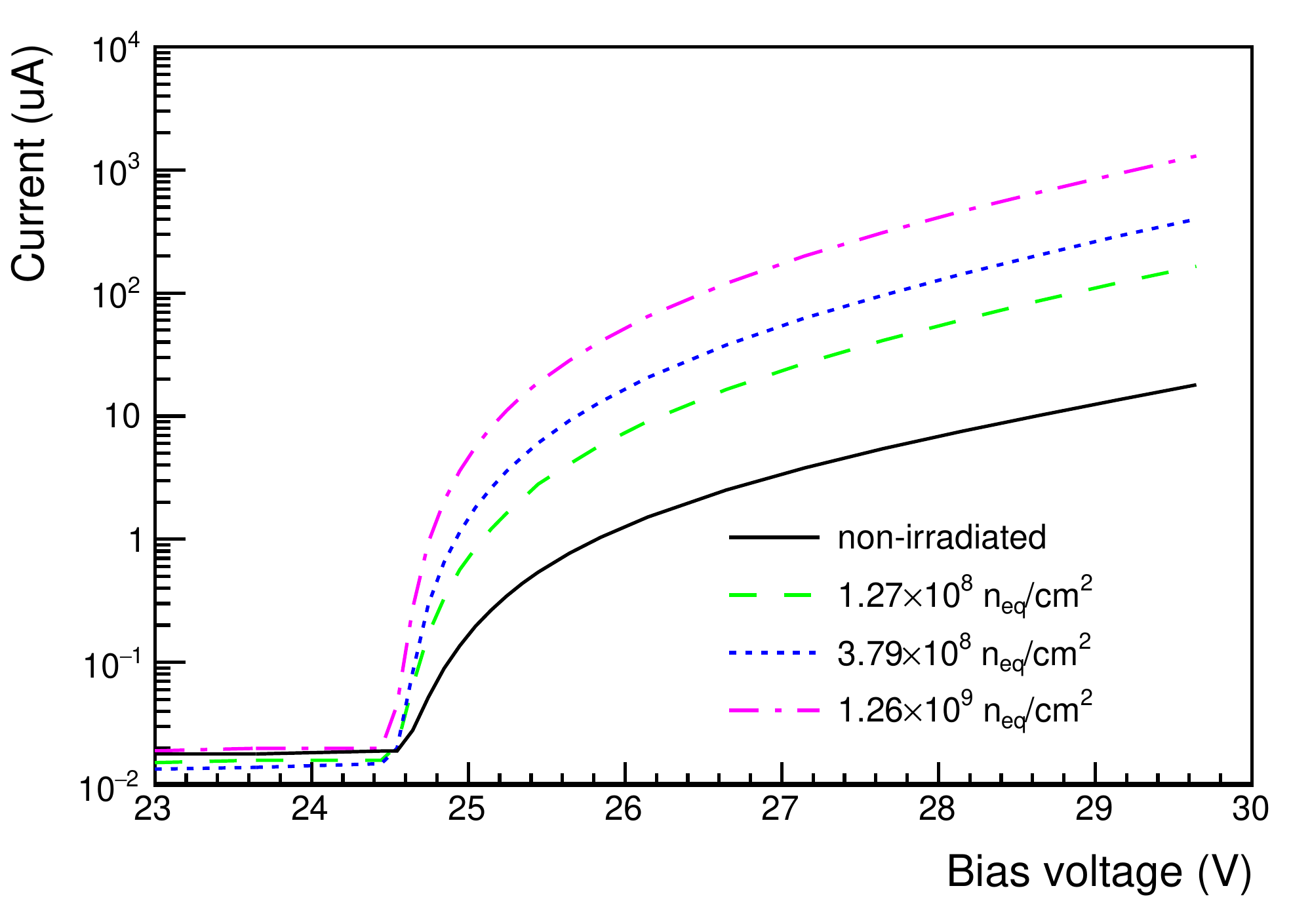}
    \caption{Current-voltage characteristics of the $2\times2$ SiPM arrays (1.47~cm$^2$) from \set{2} irradiated to different fluence levels. Measured 86 days after the irradiation. $T= 21^\circ$C.  }
    \label{fig:iv_set2}
  \end{center}
\end{figure}

A more consistent picture was observed for the \set{2} SiPMs (Figure~\ref{fig:iv_set2}). Although they were exposed to different proton fluences, all four SiPMs arrays showed similar currents below the breakdown voltage. It should be noted that there was a leakage current of 10--15~nA through the adapter PCB, which dominated the current below the breakdown voltage for the \set{2}  arrays.

\begin{figure}
  \begin{center}
    \includegraphics[width=0.8\textwidth]{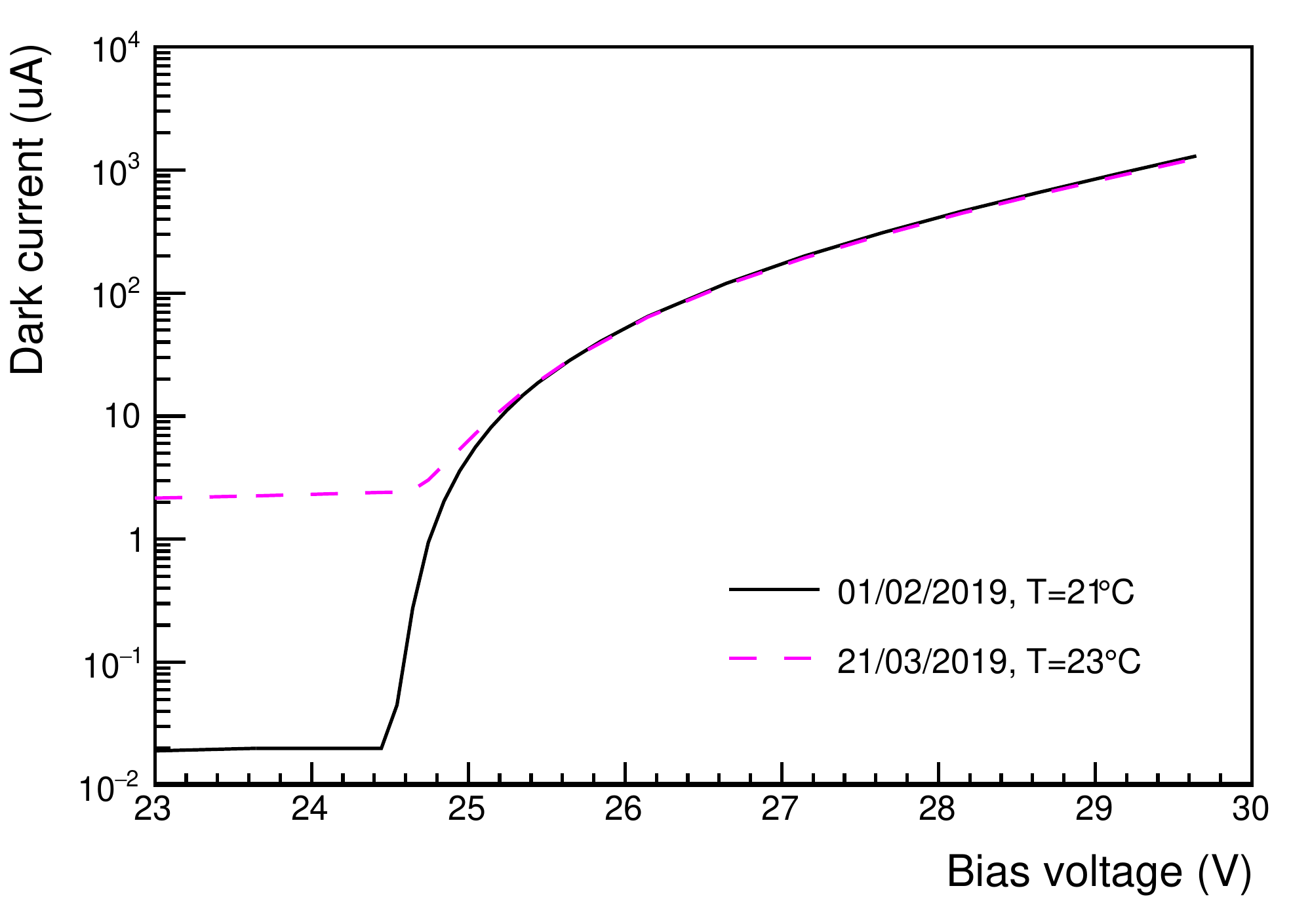}
    \caption{Current-voltage curve of Array-4 from \set{2} before and after the incident.
    }
    \label{fig:damaged_sipm}
  \end{center}
\end{figure}

The relatively large current below the breakdown voltage observed in three \set{1} SiPM arrays might have been caused by possible mechanical damage sustained by these sensors in previous studies. A similar effect was later observed in the $2\times2$ SiPM array from \set{2} irradiated to $1.26\times10^9$~\ncm (Array-4). This array was accidentally damaged when being removed from the socket adapter, which resulted in several chips along one edge of the SiPM glass cover and probably some damage to the SiPM surface. The array exhibited a large leakage current below the breakdown voltage after this incident (Figure~\ref{fig:damaged_sipm}).

In an overvoltage range of 0.3\,V $< V_\mathrm{ov} <$ 5\,V, the current-voltage curves of all irradiated $2\times2$ SiPM arrays (measured 86-87 days after irradiation) were found to be well approximated by equation:
\begin{equation}
  \label{eq:iv}
  I = I_0 + k\times F_\mathrm{eq}\times A \times (V_\mathrm{ov} ^2+0.04 V_\mathrm{ov} ^4),
\end{equation}
where $I_0$ is the SiPM dark current just below the breakdown voltage, $F_\mathrm{eq}$ is the 1~MeV neutron equivalent fluence, $A = 1.47$~cm$^2$ is the active area of the $2\times2$ SiPM array and $V_\mathrm{ov} = V_\mathrm{bias} - V_\mathrm{br}$ is the overvoltage in volts. The breakdown voltage $V_\mathrm{br}$ in Equation~(\ref{eq:iv}) was a parameter of the fit, varying for different curves from 24.50 to 24.55 V.
The factor $k$ for different curves was found to be in a range of $1.2\times 10^{-8}$ to $1.6\times 10^{-8}$~uA/n$_\mathrm{eq}$, with an average value of $1.3\times 10^{-8}$~uA/n$_\mathrm{eq}$.

\subsection{Response to gamma rays for the $25\times25\times10$ mm$^3$ CeBr3 crystal with the 16-SiPM array (\set{1})}
\label{sec:gamma_large}

\begin{figure}
  \begin{center}
    \includegraphics[width=0.45\textwidth]{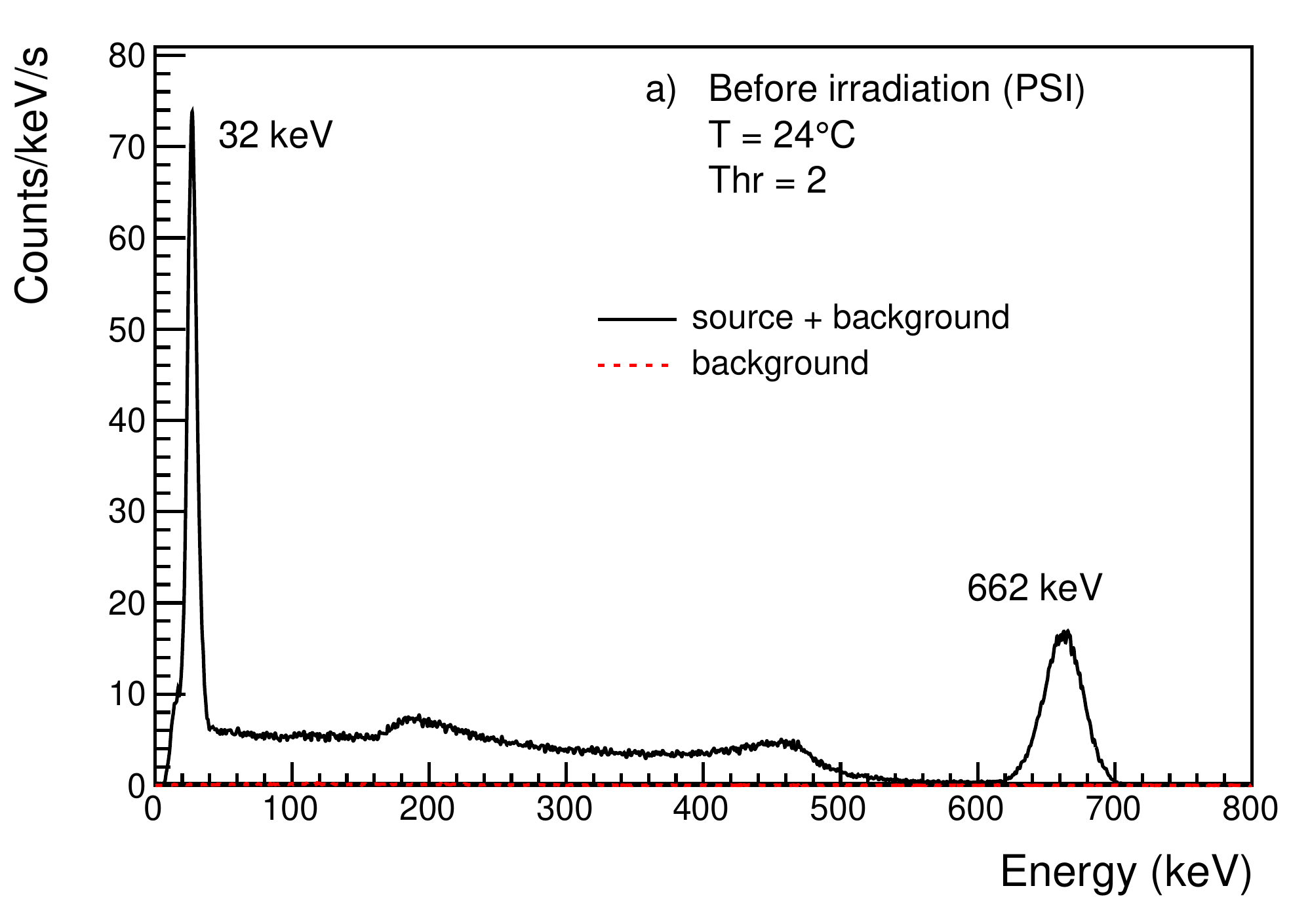}
    \includegraphics[width=0.45\textwidth]{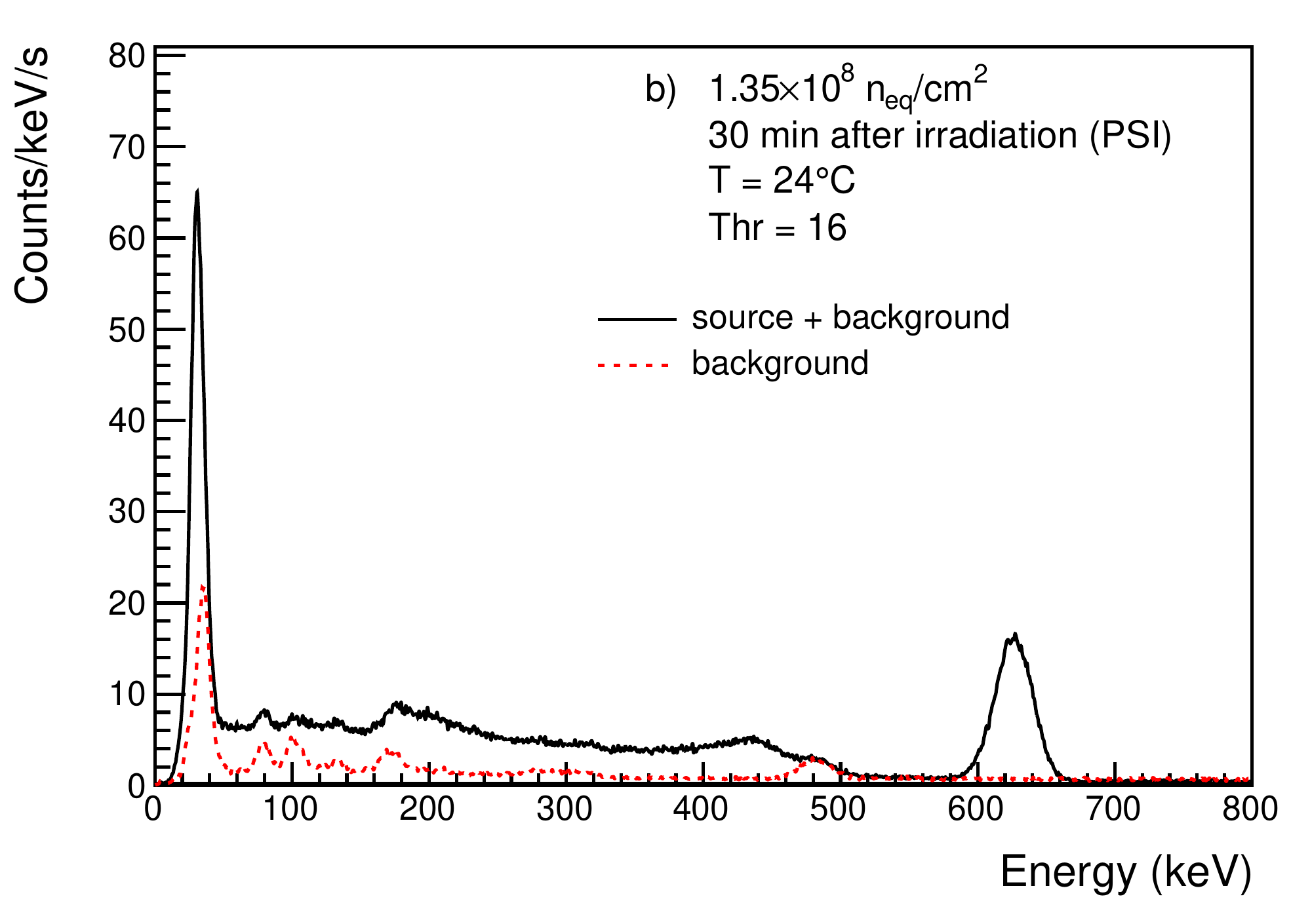}
    \includegraphics[width=0.45\textwidth]{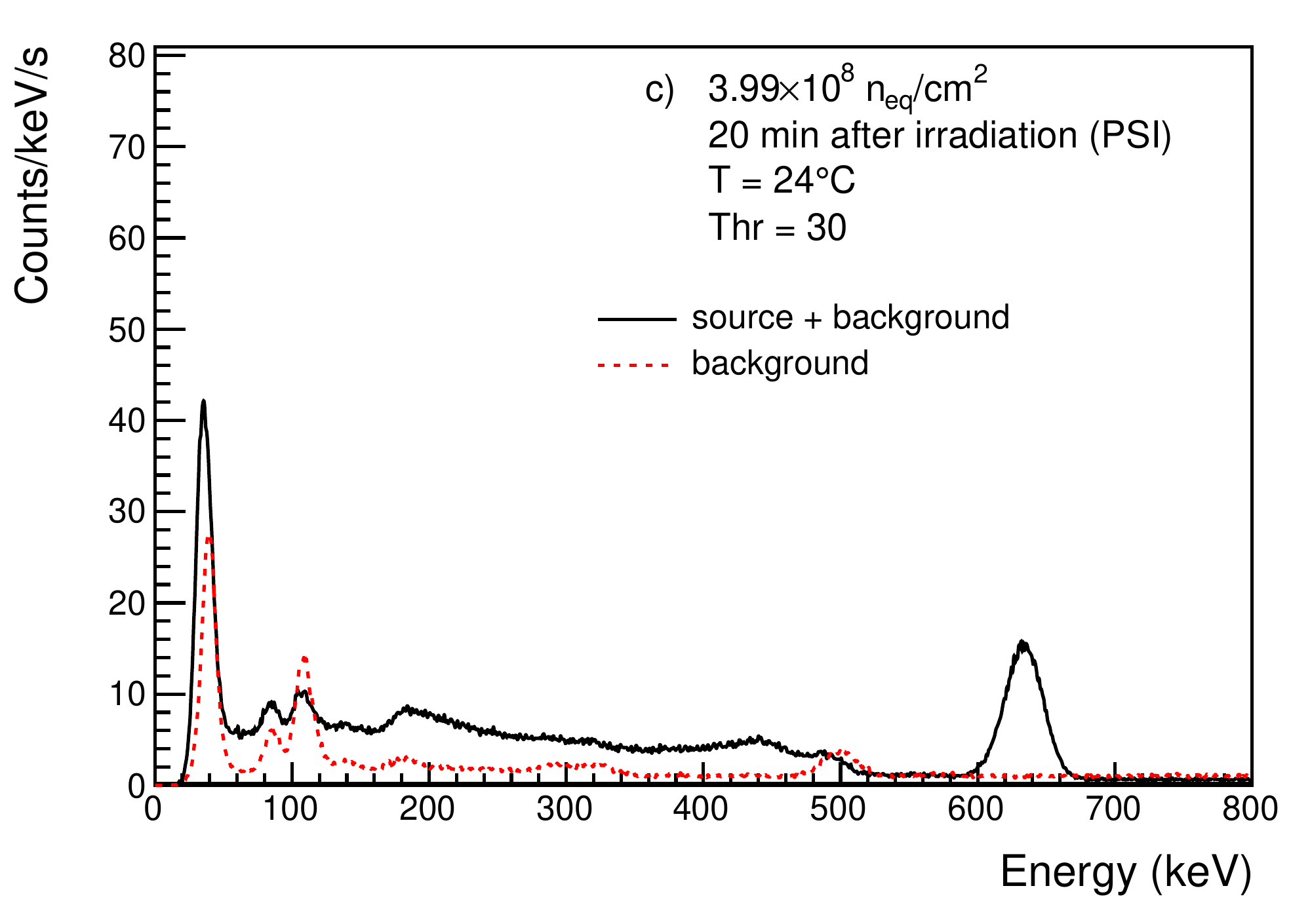}
    \includegraphics[width=0.45\textwidth]{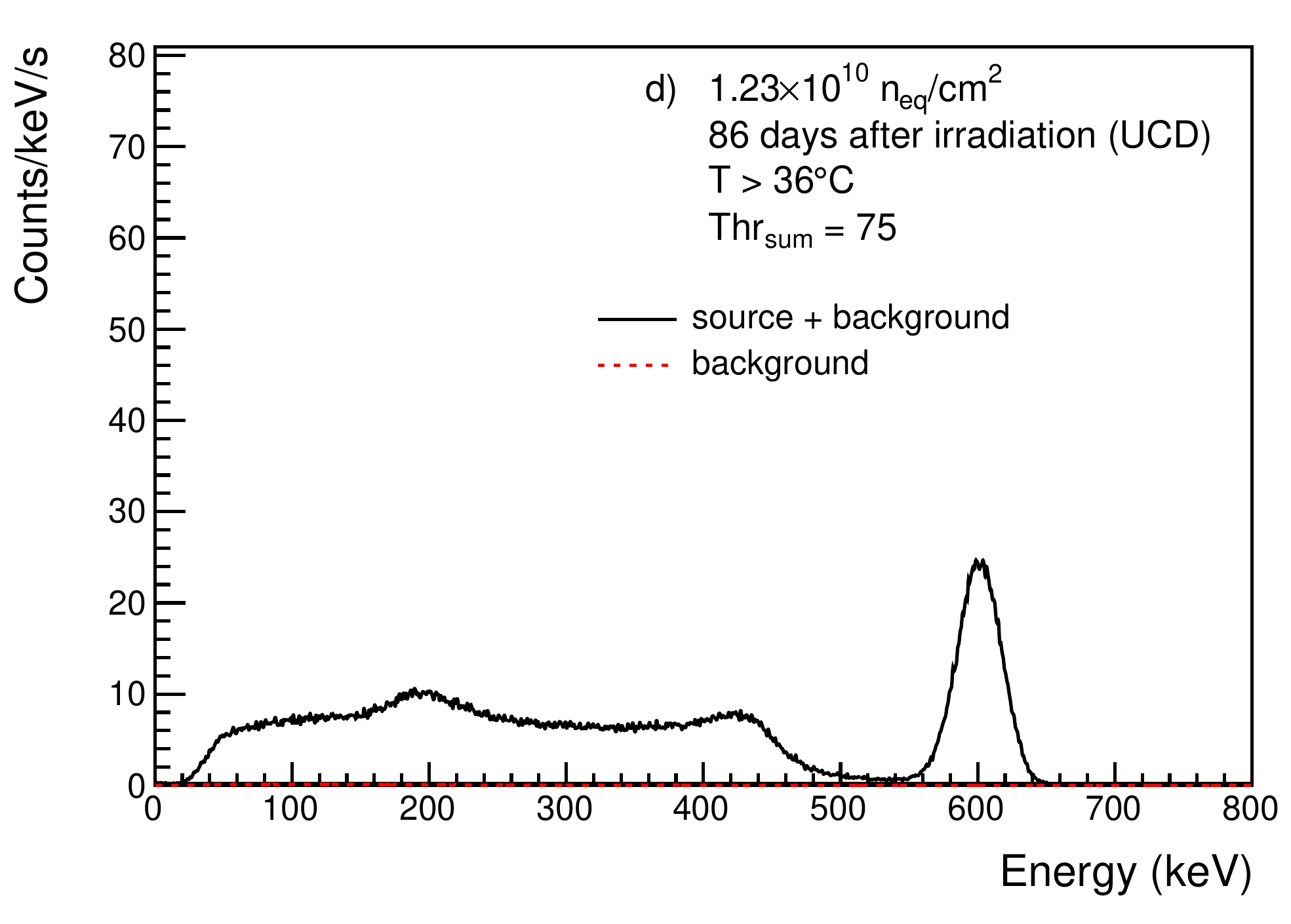}
    \caption{Spectra of $^{137}$Cs measured with the 16-SiPM (5.9~cm$^2$) CeBr3 detector before the irradiation and after several exposures to different cumulative fluences. The background spectra were acquired 3 to 5 minutes before the measurements with the $^{137}$Cs source. Before the irradiation and in the UCD measurements performed 86 days after the irradiation, the background was negligible and is not visible in the plots. The same conversion factor of 4.574 ADC channels/keV was used for all curves. The gamma-ray sources used in PSI and UCD had different levels of activity (around 5--10 uCi).}
    \label{fig:spectrum1_fl0_cs137}
  \end{center}
\end{figure}

Results of the gamma-ray spectral measurements with the large 16-SiPM CeBr3 detector are shown in Figure~\ref{fig:spectrum1_fl0_cs137} for a $^{137}$Cs source. Before the detector irradiation (Figure~\ref{fig:spectrum1_fl0_cs137}a), the measurements were performed using a trigger threshold $\mathrm{Thr}=2$ (register code in SIPHRA, the value in units of current is not known). The ADC channels were converted to keV using a scale factor of 4.574 channels/keV obtained from the position of the 662 keV gamma-ray line. Apart from the 662~keV line, the acquired spectrum showed a well-resolved peak for 32 keV X-rays emitted by the source.

After the first and second proton exposure, the trigger threshold had to be increased to 16 and 30, respectively, to account for the increased SiPM current and prevent triggers generated from the SiPM noise. The measured spectra (Figures~\ref{fig:spectrum1_fl0_cs137}b and~\ref{fig:spectrum1_fl0_cs137}c) had strong background from the detector radioactivity induced by the proton radiation in the CeBr3 scintillator and other materials.
For some energies, the background spectrum had more counts than the spectrum acquired with the source. This is because the source spectrum was acquired about 5 minutes later than the background spectrum and the level of detector radioactivity reduced during that time. For this reason it was not possible to accurately estimate the background component in the acquired source spectra and it was not subtracted.

After the last exposure ($1.23\times10^{10}$~\ncm{}), the SiPM dark current was too large for the maximum available trigger threshold of 255 and the acquired pulse-height spectrum was completely saturated by noise events. No useful spectra were measured in PSI after this exposure.

Further spectral measurements were conducted in UCD 86 days after the detector irradiation (Figure~\ref{fig:spectrum1_fl0_cs137}d). Although the SiPM current reduced by a factor of two, it was still too large for the maximum available trigger threshold of 255. To overcome this problem, triggering was performed using the summing channel in SIPHRA. The summed input current in the summing channel is divided by a factor of 16, which allowed triggering to be performed with a threshold of 75. This approach was not used in PSI.
These measurements were also complicated by SiPM heating due to the large SiPM current. After switching on the bias voltage, the SiPM temperature and current began rising fast. The gamma-ray measurements were performed after the SiPM current had stabilised at 35~mA. The SiPM temperature increased to at least $36^\circ$C (the value measured on the socket adapter PCB). The gamma-ray background rate in the UCD measurements was negligible (20 cps), as the radioactivity of the detector greatly reduced in the 86 days after the irradiation.

Apart from rejecting false events generated by the detector noise, the trigger threshold suppresses detection of low-energy gamma rays and thereby introduces a gradual low-energy cut-off in measured spectra. As the trigger threshold was raised to accommodate the increased noise of the irradiated SiPMs, the spectral cut-off increased from about 12~keV for the non-irradiated detector to about 40~keV for the detector irradiated to $1.23\times10^{10}$~\ncm{} (Figure~\ref{fig:spectrum1_fl0_cs137}). The detection efficiency of the 32~keV X-rays was greatly reduced after the detector irradiation to $3.99\times10^8$~\ncm{}, and the line was completely removed by the trigger threshold after $1.23\times10^{10}$~\ncm{}.

\begin{figure}
  \begin{center}
    \includegraphics[width=0.8\textwidth]{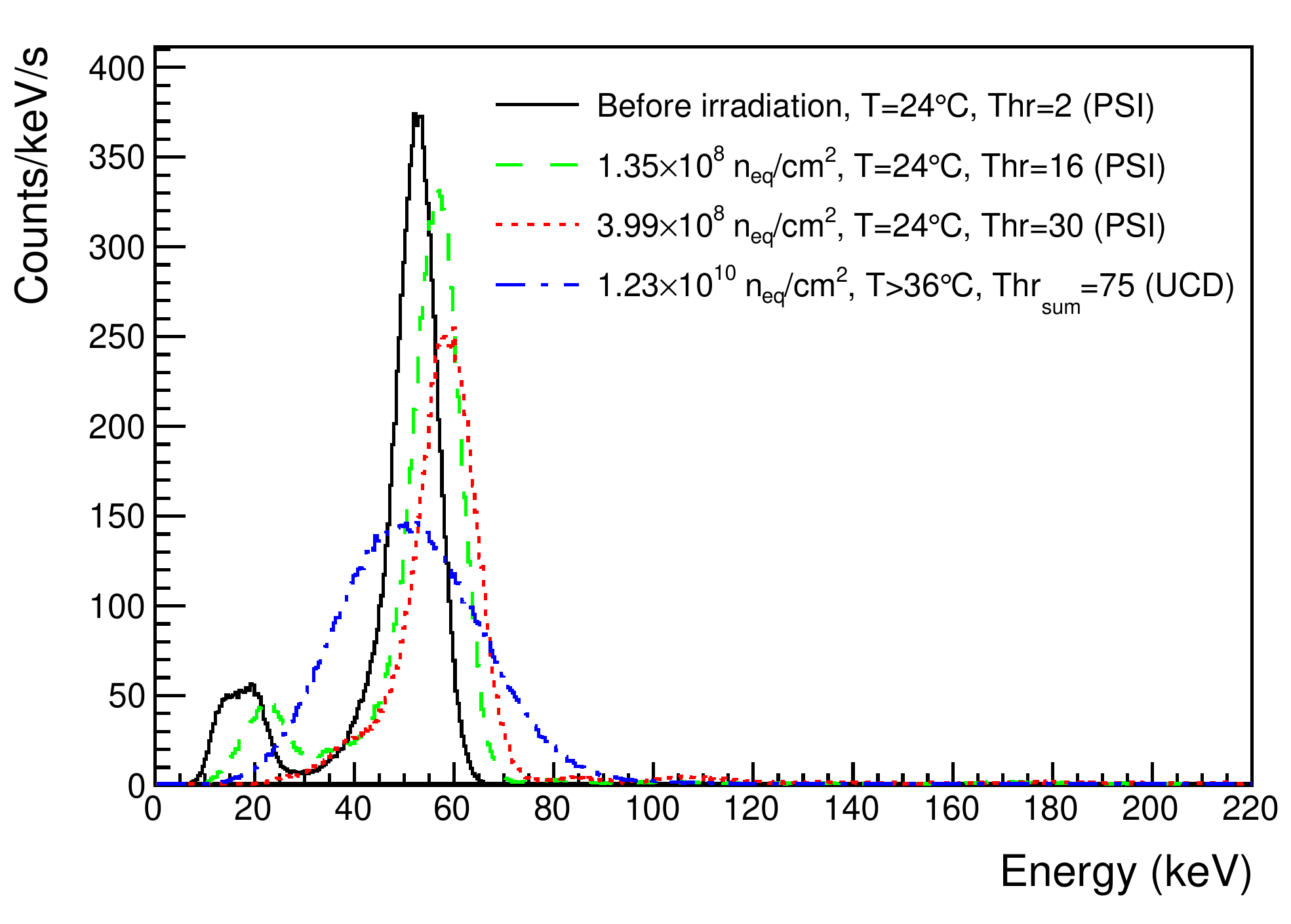}
    \caption{$^{241}$Am spectrum measured with the 16-SiPM (5.9~cm$^2$) CeBr3 detector after irradiation to different cumulative fluence levels. The same conversion factor of 4.574 ADC channels/keV was used for all curves. Two different $^{241}$Am sources were used in PSI and UCD, each with an activity of around 10~uCi.}
    \label{fig:spectrum1_am241}
  \end{center}
\end{figure}

After each exposure, a spectrum of a $^{241}$Am source was measured several minutes after recording a $^{137}$Cs spectrum (Figure~\ref{fig:spectrum1_am241}). Compared to the $^{137}$Cs source, the differential count rates with  the $^{241}$Am source were much higher and the contribution of the background radioactivity was not that important. The spectrum obtained with the non-irradiated detector showed a strong peak from 59.5 keV gamma rays and a smaller peak from multiple unresolved lines in a range of 10-26~keV. The smaller low-energy peak disappeared as the trigger threshold was increased after the irradiation of the detector. However, the detector was still able to detect the 59.5 keV gamma rays after the final exposure (cumulative fluence of $1.23\times10^{10}$~\ncm{}) and the 86-day recovery period.

The measured positions and widths of the 662 keV and 59.5~keV gamma-ray lines are summarised in Table~\ref{table:lines}.
The detector energy resolution was deteriorated by the SiPM noise increasing after each detector irradiation, but the effect was mainly important at low energies. In addition, significant shifts in the positions of the spectral peaks were observed after each proton exposure. These shifts can mostly be attributed to changes in the SiPM temperature and changes in the trigger threshold (and the change to the summing trigger channel after the last irradiation) which has a considerable and non-linear effect on pulse-height measurements with SIPHRA (Section~\ref{sec:siphra}). However, the downward shift in the position of the 662 keV peak after the first irradiation ($1.35\times10^8$~\ncm{}) is not understood, as an increase of the trigger threshold was found to have the opposite effect. In any case, it was not possible to accurately evaluate the effect of proton irradiation on the scale of SiPM signals with the measurement set-up used in this study. If any effect was present, it did no exceed 10\%.

\begin{table}[h]
  \caption{Mean energy and full width at half maximum of the 59.5 and 662 keV gamma-ray lines measured by the 16-SiPM (5.9~cm$^2$) CeBr3 detector using a bias voltage of 28.15~V. The same conversion factor of 4.574 ADC channels per keV was used in all measurements.}
  \label{table:lines}
  \begin{center}
  \begin{tabular}{c c c c c c c}
    \hline
     Fluence & Time after   & Temperature & \multicolumn{2}{c}{59.5 keV} & \multicolumn{2}{c}{662 keV}  \\
     (\ncm{})     & irradiation  & ($^\circ$C) & Mean & FWHM     & Mean & FWHM   \\
    \hline
 $ 0 $             &     -         & 24 & 52.5 keV & 18\% & 662 keV & 5.1\%    \\
 $1.35\times10^8$   & 30 min       & 24 & 56.4 keV & 19\% & 626 keV & 5.1\%     \\
 $3.99\times10^8$   & 20 min       & 24 & 58.6 keV & 22\% & 633 keV & 5.1\%    \\
 $1.23\times10^{10}$   & 86 days      & 36$^*$ & 51.6 keV & 64\% & 601 keV & 6.2\%     \\

      \hline
  \end{tabular}
  \end{center}
    \footnotesize * The temperature measured on the socket adapter board. The SiPM temperature was probably higher.
 \end{table}

The observed non-linearity of the detector is caused by the non-linearity of the CeBr3 response~\cite{quarati_cebr3}, non-linearity of the SiPMs and the effects of the readout system. The non-linearity of the SiPMs in the used signal range is expected to be relatively small: assuming 40 scintillation photons are collected by the SiPM array per 1 keV deposited in the CeBr3 crystal, the signal from a fully absorbed 662~keV gamma ray corresponds to about 11000 fired microcells, or about 3\% of the total number of microcells in the 16-SiPM array. This results in SiPM signal non-linearity of about 1.5\%, although the effect may be slightly underestimated due to statistical fluctuations and non-uniformity of light distribution across the SiPM array.

\subsection{Response to gamma rays for the $10\times 10\times 10$ mm$^3$ CeBr3 crystal with a 4-SiPM array (\set{1} / \set{2})}
\label{sec:gamma_small}
Gamma-ray measurements using the small CeBr3 scintillator coupled to one $2\times2$ SiPM array were conducted in UCD on 20 March 2019, 134 days after the SiPM irradiation. The measurements were performed for all four \set{2} SiPM arrays individually irradiated to different fluence levels, as well as for one SiPM array (Array-1) from \set{1} irradiated to $1.23\times10^{10}$~\ncm{} as part of the large detector.

The measurements were performed at a temperature of 22$^\circ$C. For the SiPM array irradiated to $1.23\times10^{10}$~\ncm, the SiPM temperature was significantly higher because of the large dark current (26$^\circ$C was measured on the socket adapter board). However, it was not as high as it was in the similar measurements performed with the 16-SiPM detector (Section~\ref{sec:gamma_large}), probably due to the absence of detector housing and less heat generated with fewer SiPMs (Figure~\ref{fig:measurement_setup_single}). As a result of the lower temperature, the SiPM dark current in these measurements was significantly smaller (5.5~mA) and a trigger threshold of 235 was sufficient to prevent triggering from the SiPM noise.

The detector response to 662 keV gamma rays from $^{137}$Cs measured with different SiPM arrays varied by several percent because of small variations in coupling and positioning of the scintillator. To account for this difference, an individual ADC-channel-to-keV conversion factor was calculated for each SiPM array using the position of the 662 keV peak. This is different from the measurements discussed in Section~\ref{sec:gamma_large}, where all measurements were performed using a single detector and a single calibration factor was used to observe the effects of the irradiation on the detector response.

\begin{figure}
  \begin{center}
    \includegraphics[width=0.8\textwidth]{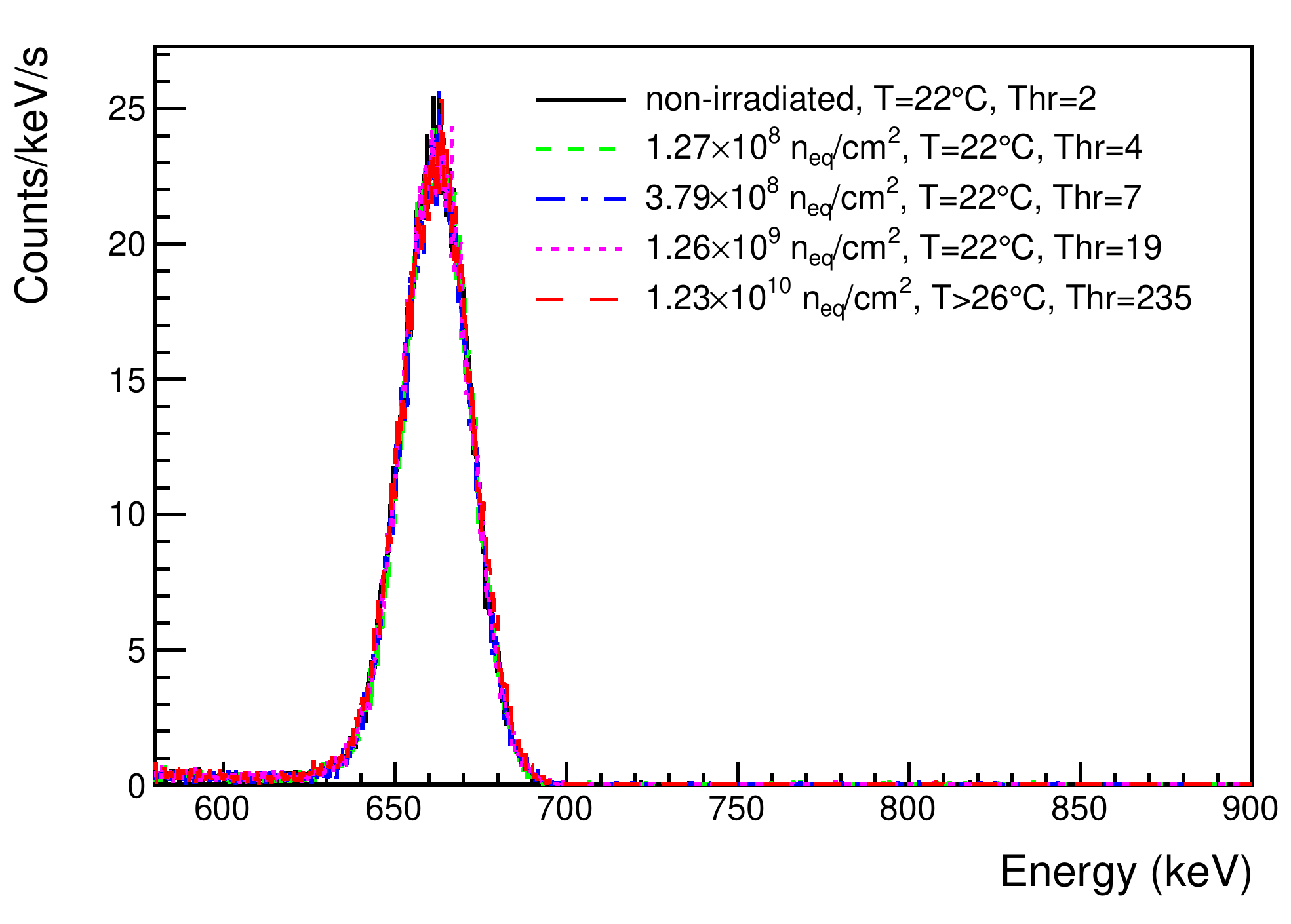}
    \caption{662 keV line in the $^{137}$Cs spectrum measured with a 4-SiPM (1.47~cm$^2$) CeBr3 detector.  }
    \label{fig:spectrum_small_cs137b}
  \end{center}
\end{figure}

\begin{figure}
  \begin{center}
    \includegraphics[width=0.8\textwidth]{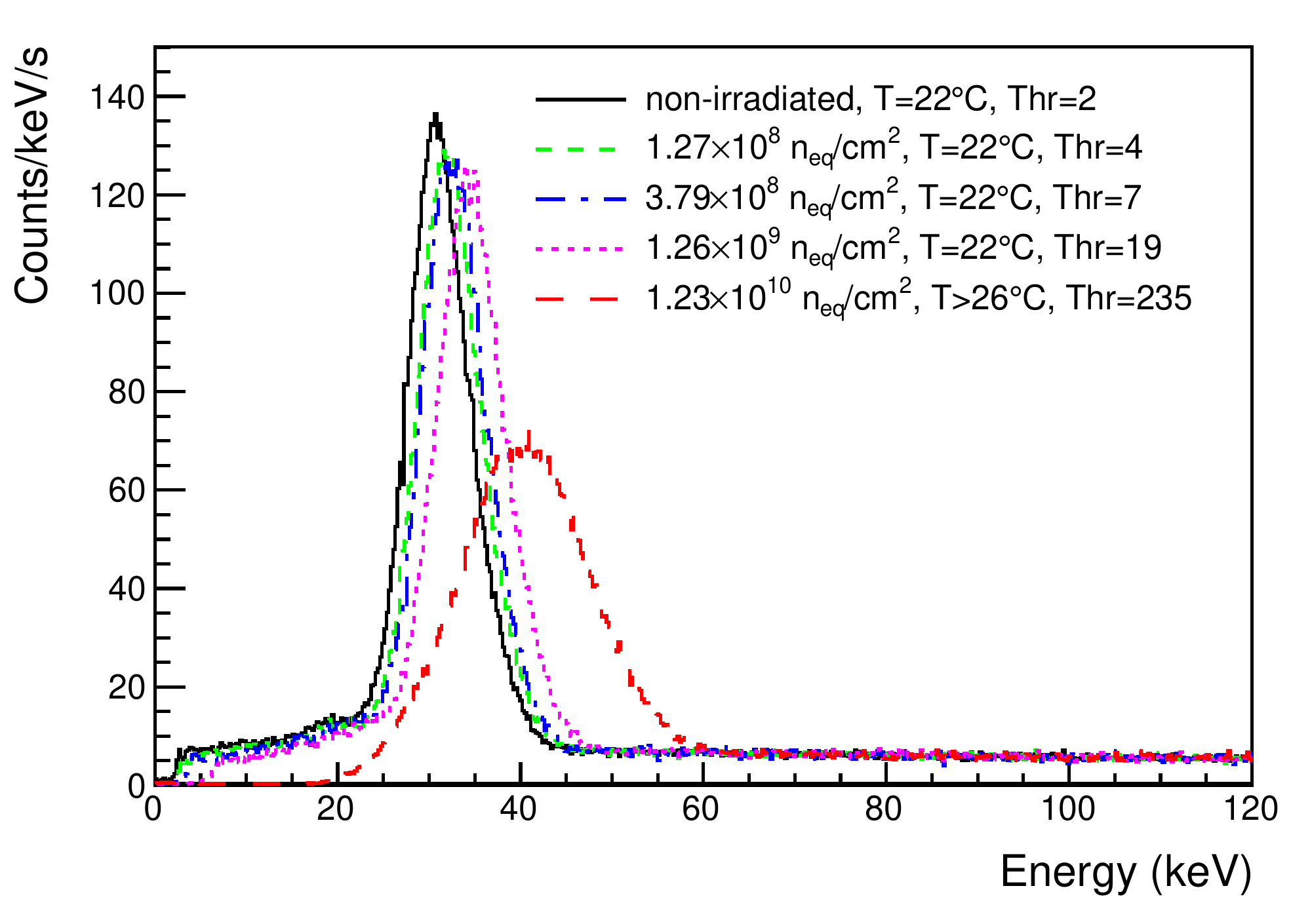}
    \caption{(colour online) 32 keV X-ray line in the $^{137}$Cs spectrum measured with a 4-SiPM (1.47~cm$^2$) CeBr3 detector. }
    \label{fig:spectrum_small_cs137a}
  \end{center}
\end{figure}

\begin{figure}
  \begin{center}
    \includegraphics[width=0.8\textwidth]{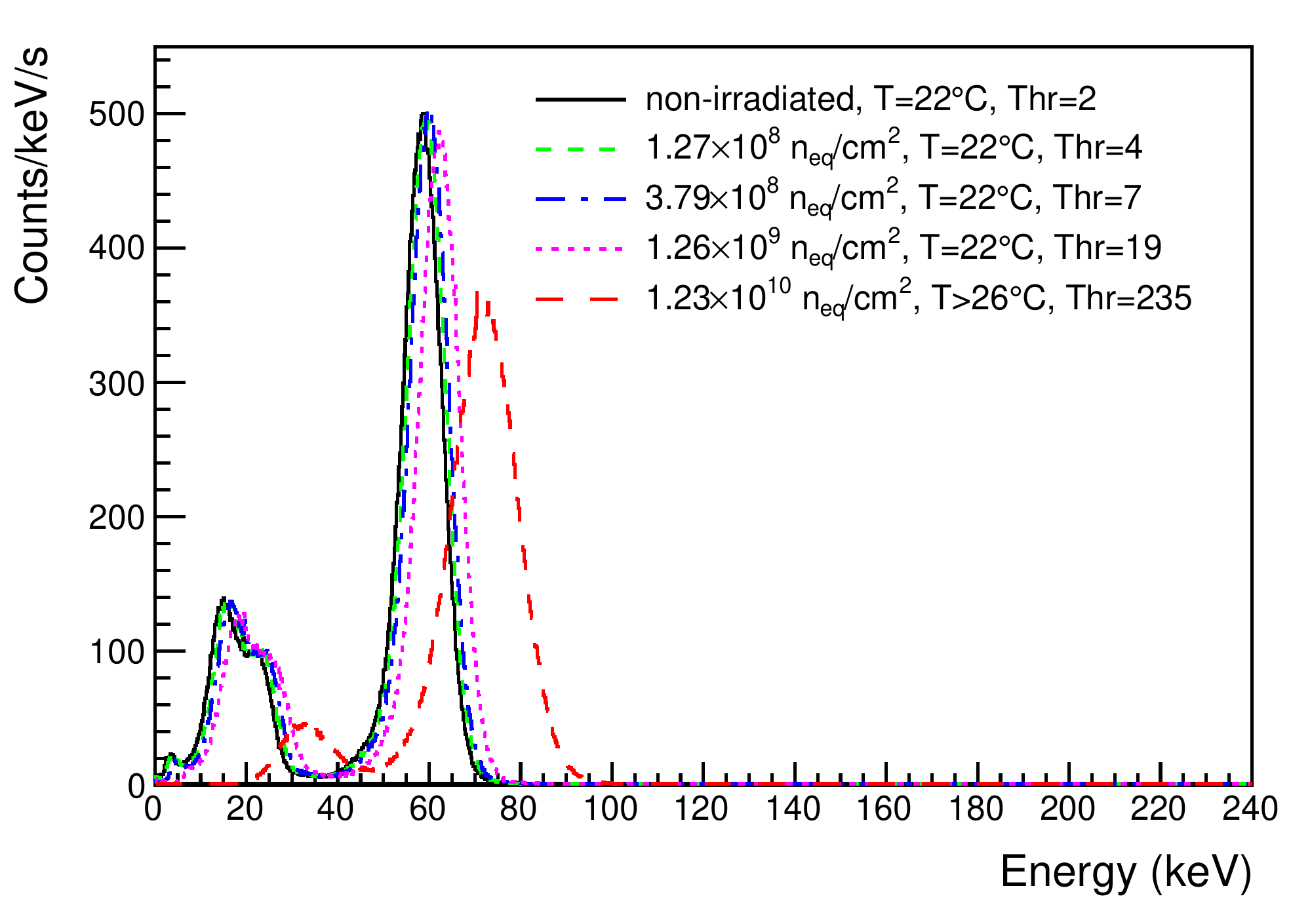}
    \caption{(colour online) $^{241}$Am spectrum measured with a 4-SiPM (1.47~cm$^2$) CeBr3 detector. }
    \label{fig:spectrum_small_am241}
  \end{center}
\end{figure}

After scale calibration of each detector at 662~keV, the 662 keV peaks in the spectra obtained with $2\times2$ SiPM arrays exposed to different fluences were virtually identical (Figure~\ref{fig:spectrum_small_cs137b}).
However, the 32 keV X-ray peak and 59.5 keV gamma-ray peak in the $^{241}$Am spectrum were affected by the SiPM irradiation as shown in Figures~\ref{fig:spectrum_small_cs137a} and~\ref{fig:spectrum_small_am241}. The shift in the position of the spectral peaks after the irradiation is again explained by the effect of the delayed signal sampling in SIPHRA caused by the increased trigger threshold.

Due to the smaller number of SiPMs, the 4-SiPM detector had less noise compared to the 16-SiPM detector and therefore showed better spectral performance. Unlike the larger detector, it was able to detect 32 keV X-rays using the SiPMs irradiated to $1.23\times10^{10}$~\ncm{}. In addition, triggering on low-energy signals was more efficient as it was performed using the total detector signal (in one SIPHRA input channel) in contrast to triggering in four separate SIPHRA channels with partial detector signals in the case of the larger detector. Measured positions and widths of the X-ray and gamma-ray lines are summarised in Table~\ref{table:lines2}.

\begin{table}[h]
  \caption{Mean energy and full width at half maximum of the gamma-ray lines measured with the 4-SiPM (1.47~cm$^2$) CeBr3 detector 134 day after the SiPM irradiation. For each fluence level, a different SiPM array was used and the detector was recalibrated using the 662~keV line.}
  \label{table:lines2}
  {
  \resizebox{\textwidth}{!}{
  \begin{tabular}{c c c c c c c c c}
    \hline
     Fluence  & Temperature & \multicolumn{2}{c}{32 keV}  & \multicolumn{2}{c}{59.5 keV} & \multicolumn{2}{c}{662 keV} & Calibration \\
     \ncm{} & ($^\circ$C) & Mean & FWHM & Mean & FWHM & Mean & FWHM & (ADC ch/keV) \\
    \hline
 $ 0$               & 22 & 31.1 keV & 26\% & 58.4 keV & 17\% & 662 keV & 3.5\% &  4.178 \\
 $1.27\times10^8$    & 22 & 32.3 keV & 26\% & 59.1 keV & 17\% & 662 keV & 3.5\% & 4.248   \\
 $3.79\times10^8$    & 22 & 32.8 keV & 26\% & 59.9 keV & 17\% & 662 keV & 3.5\%  & 4.127  \\
 $1.26\times10^9$    & 22 & 34.6 keV & 26\% & 62.1 keV & 17\% & 662 keV & 3.5\%  & 4.154  \\
 $1.23\times10^{10}$ &  26$^*$ & 40.6 keV & 40\% & 72.1 keV & 23\% & 662 keV & 3.6\% & 3.884 \\
    \hline
  \end{tabular}
  }
  }
  \footnotesize * The temperature measured on the socket adapter board. The SiPM temperature was probably higher.

 \end{table}

\subsection{SiPM dark noise}
\label{sec:noise}
\begin{figure}
  \begin{center}
    \includegraphics[width=0.45\textwidth]{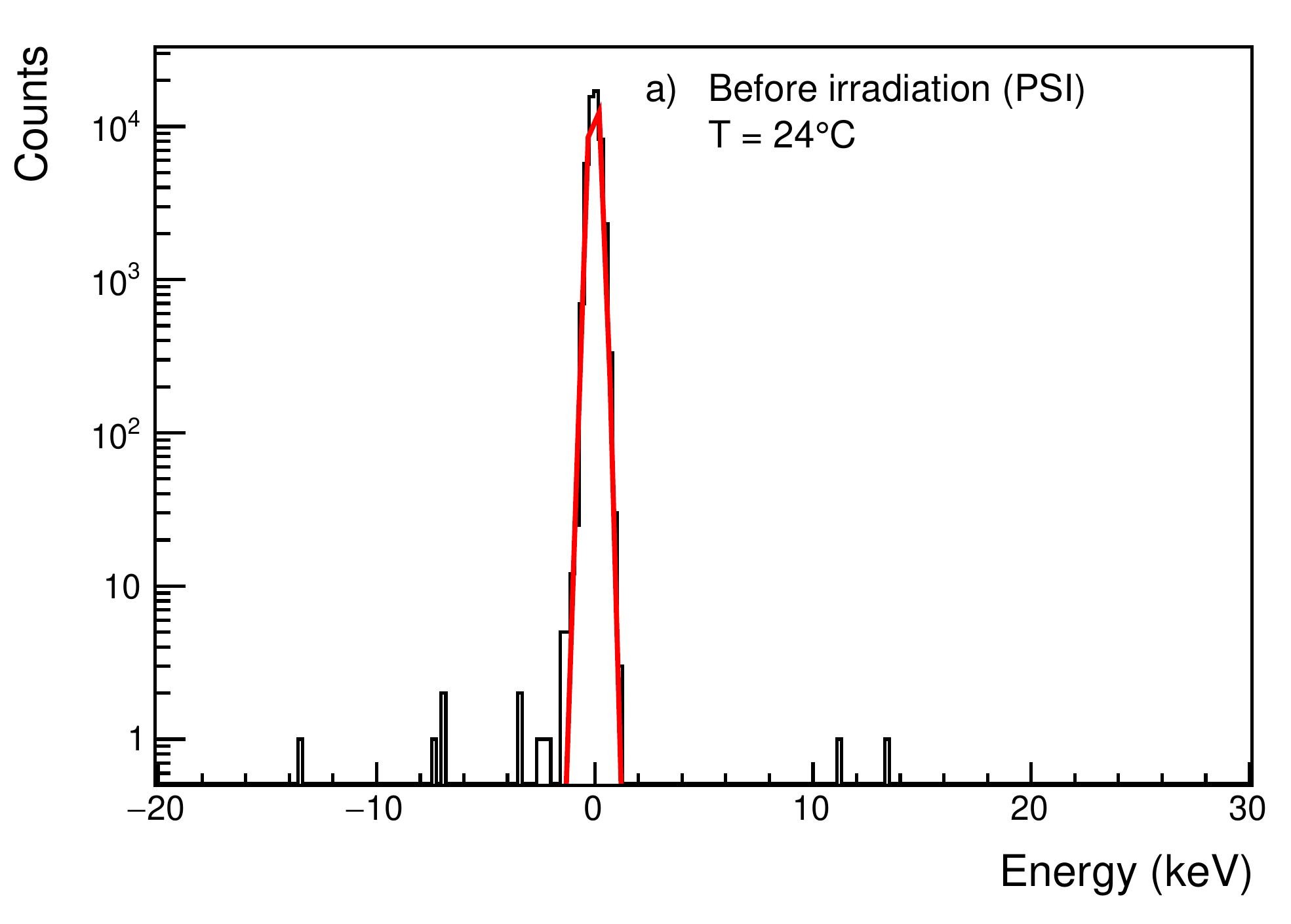}
    \includegraphics[width=0.45\textwidth]{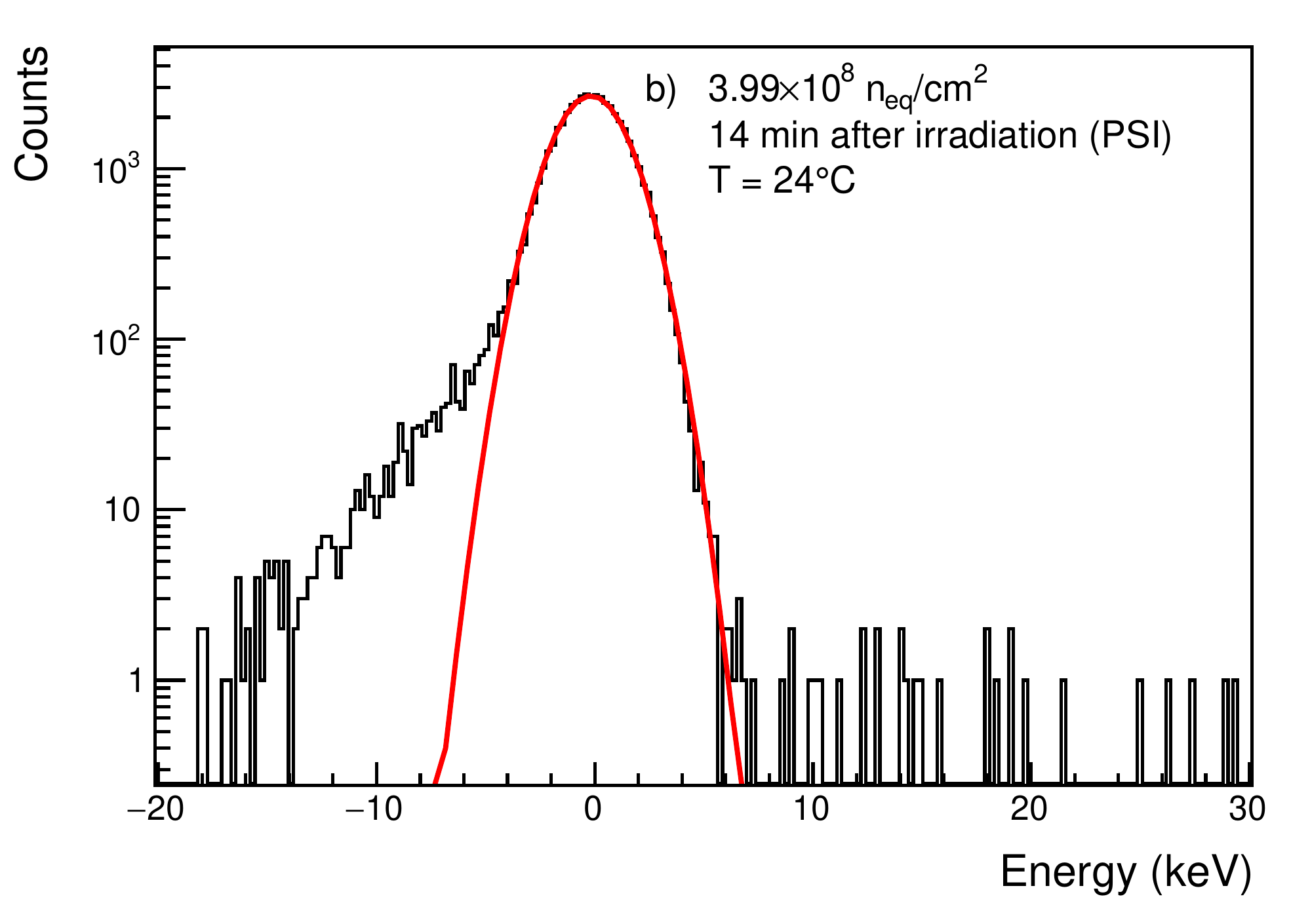}
    \includegraphics[width=0.45\textwidth]{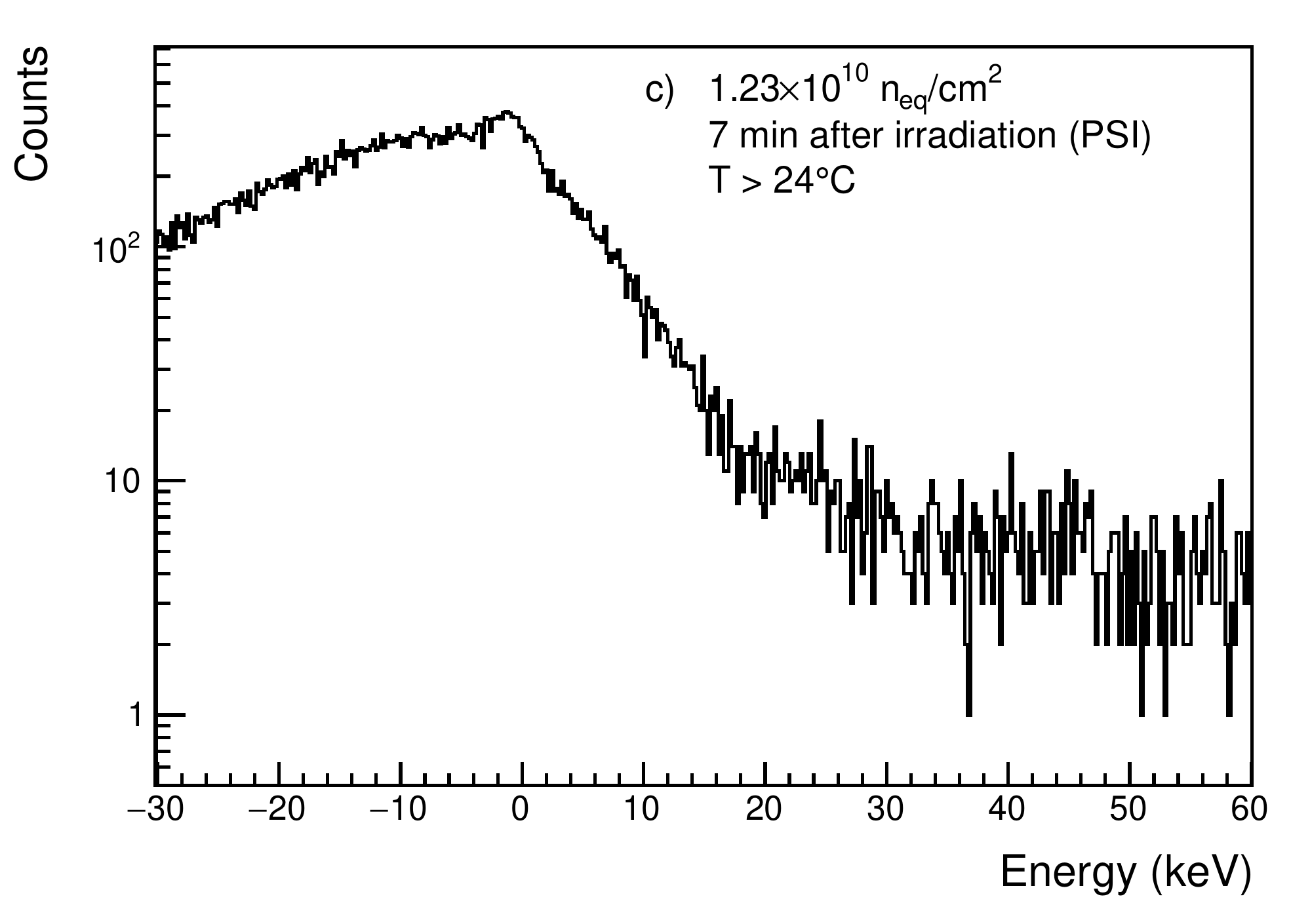}
    \includegraphics[width=0.45\textwidth]{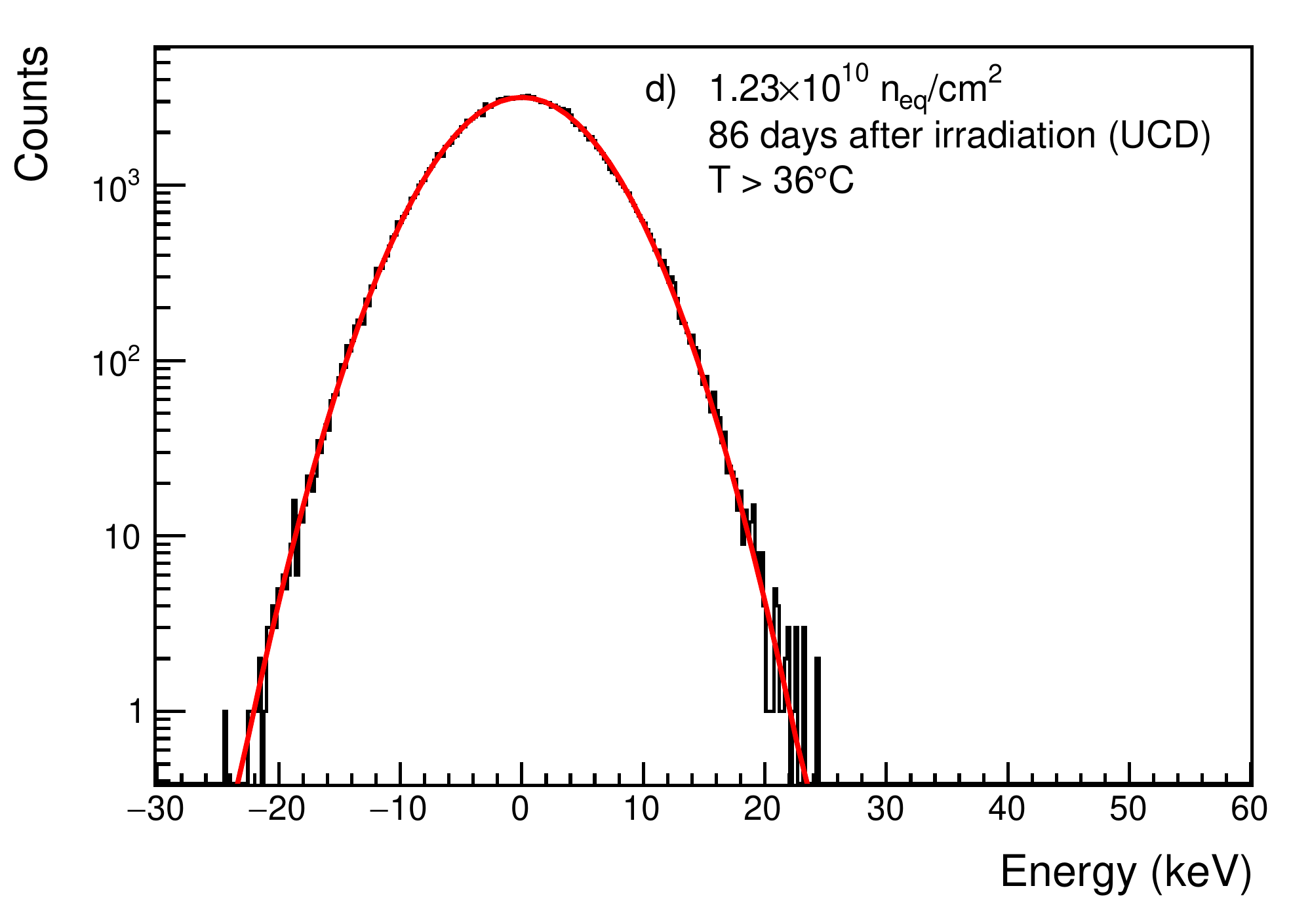}
    \caption{Noise distribution in one detector channel (one $2\times2$ SiPM array from \set{1}) before and after detector irradiation. The smooth curve shows a Gaussian function fitted to the spectrum.}
    \label{fig:noise100}
  \end{center}
\end{figure}

SiPM dark noise was measured with the same experimental set-up as  used in  the gamma-ray measurements. However, instead of using the threshold trigger, SiPM dark signals were randomly sampled by SIPHRA using an external trigger. Measurements for \set{1} were performed with the CeBr3 scintillator attached to the SiPM array which emitted some scintillation light due to background radiation. Noise spectra measured after different proton exposures are shown in Figure~\ref{fig:noise100}. ADC channels were converted to keV using the scale factor of 4.574 ADC channels per keV obtained for the CeBr3 detector using 662~keV gamma rays (Section~\ref{sec:gamma_large}).  The distributions show randomly sampled noise which is directly added to any energy measurement by the gamma-ray detector. The average value of SiPM noise with SIPHRA readout is zero, therefore the noise does not contribute to the average measured energy but adds statistical fluctuations, contributing to the widths of spectral lines. The noise distributions should not be confused with spectra of noise events which are acquired when noise pulses exceed the trigger threshold.  The measured distributions contain all possible values of noise including peak values.

The noise distribution measured before the detector irradiation (Figure~\ref{fig:noise100}a)
is well described by a Gaussian distribution. Some off-peak counts observed in the spectrum are due to occasional overlap of randomly acquired events with background gamma rays. In the spectrum measured immediately after a proton exposure (Figure~\ref{fig:noise100}b), the scintillator radioactivity induced by protons results in noticeable tails. The tail extending to negative amplitudes is more pronounced. This is explained by the shape of the SiPM signal at the output of the SIPHRA shaper: the duration of the main pulse is about 1~$\mu$s which is followed by a long ($\sim15$~$\mu$s) over-shoot below the baseline.
Apart from the asymmetric tails caused by rare background events, the measured spectrum is defined by the SiPM noise.
The dark count rate of the SiPM array is very high, therefore any sampling of the shaped signal has contributions from many dark pulses, with some pulses making positive contributions and other (earlier) pulses making negative contributions. Due to a large number of contributions, the resulting noise distribution has a symmetric Gaussian shape.

The noise spectrum measured in PSI after the third proton exposure (Figure~\ref{fig:noise100}c)  was so badly affected by the induced radioactivity of the scintillator that it was not possible to distinguish the SiPM dark noise from pile-up of background events. When the measurements were repeated 86 days later in UCD, the detector radioactivity reduced to a negligible level and the SiPM noise was again described by a Gaussian distribution (Figure~\ref{fig:noise100}d).

The above spectra are shown for individual SIPHRA channels, each connected to a $2\times2$ SiPM array. The noise was found to be very similar for all four arrays. The noise of the entire 16-SiPM array was measured using a distribution of the sum of the digitised signals from all four channels. The sigma of a Gaussian function fitted to the noise distribution was used as a measure of the noise. The results of all measurements including \set{1} and \set{2} are summarised in Table~\ref{table:noise}.  The noise of the entire 16-SiPM array was slightly larger than 2 times the noise of one 4-SiPM array, indicating a small correlation in the noise of the four arrays.

The SiPM dark noise greatly increased after the SiPM irradiation with protons. After irradiation to a cumulative fluence of $1.23\times10^{10}$~\ncm{}, heating of the SiPMs by the large dark current significantly contributed to the increase of SiPM noise. In this case it was possible to reduce the SiPM temperature and noise by operating the SiPMs at a lower bias voltage (the detector was calibrated at each voltage using 662 keV gamma-ray line). The dark noise measured for \set{2} SiPMs in UCD was significantly lower than the dark noise measured in PSI for \set{1} SiPMs after similar radiation exposure. This difference was likely caused by partial recovery of the displacement damage during 89 days after the irradiation and a $4^\circ$C lower temperature for the UCD measurements.

\begin{table}[h]
  \caption{$1\sigma$ noise of 4 SiPMs (1.47 cm$^2$) and 16 SiPMs (5.9 cm$^2$) after proton irradiation to different fluence levels. The energy scale is for a CeBr3 gamma-ray detector. Internal SIPHRA noise of 0.69 ADC channels (measured with the SiPM bias voltage switched off) has been subtracted.}
  \label{table:noise}
  \begin{center}
    \resizebox{\textwidth}{!}{
    \begin{tabular}{c c c c c c c c c}
    \hline
    SiPM &  Fluence    & Time after   & Voltage & Temperature & \multicolumn{2}{c}{Noise (ADC channels)} & \multicolumn{2}{c}{Noise (keV)}  \\
    set  &  (\ncm{})      & irradiation  & (V)     &  ($^\circ$C) &4 SiPMs & 16 SiPMs &4 SiPMs & 16 SiPMs   \\
    \hline
 1 & $ 0$             & --       & 28.15 & 24 & 0.81 &  1.81 & 0.18 & 0.40  \\
 1 &$1.35\times10^8$   & 28 min  & 28.15 & 24 & 4.98 & 10.5 &  1.09 & 2.3     \\
 1 &$3.99\times10^8$   & 14 min  & 28.15 & 24 & 7.32 & 16.3 & 1.6 & 3.6     \\
 \hline
 1 &$1.23\times10^{10}$   & 86 days  & 28.15 & 36$^*$ & 27.4 & 59.8 & 6.0 & 13.1     \\
 1 &$1.23\times10^{10}$   & 86 days  & 26.95 & 23$^*$ & 9.73 & 20.9 & 4.0 & 8.6     \\
 1 &$1.23\times10^{10}$   & 86 days  & 26.25 & 22$^*$ & 5.21 & 11.1 & 3.6 & 7.7     \\
 \hline
 2 &$ 0$              &  --       & 28.15 & 20 & 0.77 & --  & 0.17 &  --   \\
 2 & $1.27\times10^8$  & 89 days  & 28.15 & 20 & 2.15 & --  & 0.47  &  --   \\
 2 & $3.79\times10^8$  & 89 days  & 28.15 & 20 & 3.40 & --  & 0.74 &  --    \\
 2 & $1.26\times10^9$  & 89 days  & 28.15 & 20 & 5.64 & --  & 1.24 & --    \\
      \hline
  \end{tabular}
  }
  \end{center}
  \footnotesize * The temperature measured on the socket adapter board. The real SiPM temperature was possibly higher.
\end{table}

\subsection{SiPM annealing}
As discussed in Section~\ref{sec:current}, significant recovery of the proton induced radiation damage was observed in SiPMs in 86 days after the irradiation under normal conditions: the dark current measured in UCD was about three times lower than the current measured in PSI immediately after the irradiation.

24-hour room-temperature annealing under nominal bias voltage (28.15~V) was performed for the \set{2} SiPM arrays on 21 March 2019 (132 days after the irradiation). No significant change in the SiPM dark current was observed as a result of this procedure (Table~\ref{table:rt_annealing}).

\begin{table}[h]
  \caption{Dark current of $2\times2$ SiPM arrays from \set{2} before and after 24-hour room-temperature annealing under a bias voltage of 28.15~V. All measurements were performed at 23~$^\circ$C.}
  \label{table:rt_annealing}
  \begin{center}
  \begin{tabular}{c c c c}
    \hline
    SiPM &  Fluence &  \multicolumn{2}{c}{Dark current (uA)}  \\
    array          &(\ncm{})  & before annealing  & after annealing \\
    \hline
 1 & $ 0 $  &  8.28  & 8.16  \\
 2 & $1.27\times10^8$   & 63.8   & 67.0  \\
 3 & $3.79\times10^8$   & 136   &  136 \\
 4 & $1.26\times10^9$   & 444   &  442 \\
       \hline
  \end{tabular}
  \end{center}
\end{table}

168-hour annealing (including two stages of 25 and 143 hours) at a temperature of 79$^\circ$C was performed on 11-18 April 2019 (154 days after irradiation) simultaneously for all \set{1} and \set{2} SiPM arrays except the non-irradiated Array-1 from \set{2}. To investigate a potential effect of the bias voltage, Array-3 and Array-4 from \set{1} were biased at 24~V during the annealing procedure. The bias voltage was set below the SiPM breakdown voltage to avoid a large current through the SiPMs and additional heating. All other SiPMs were unbiased.

\begin{figure}
  \begin{center}
    \includegraphics[width=0.45\textwidth]{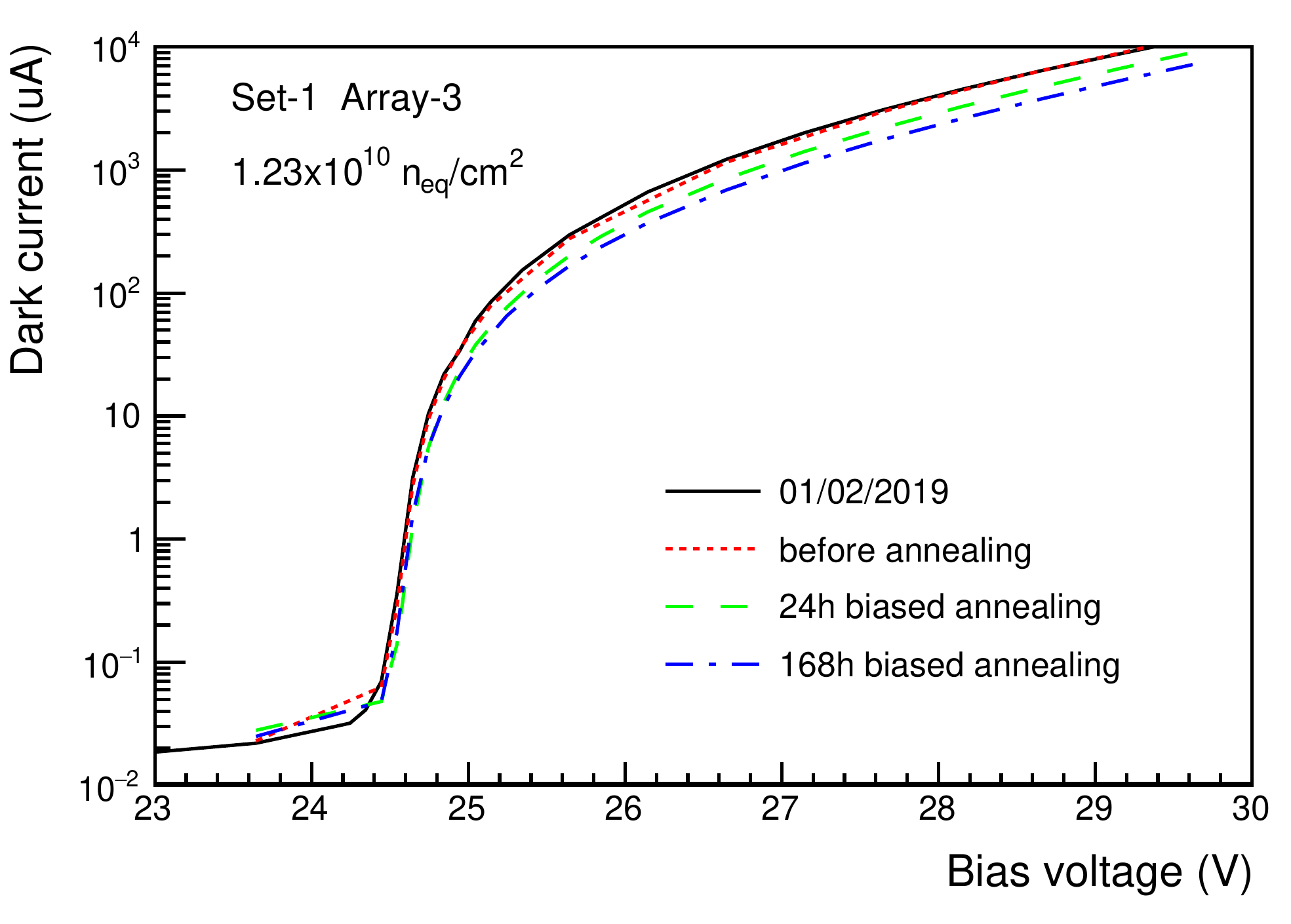}
    \includegraphics[width=0.45\textwidth]{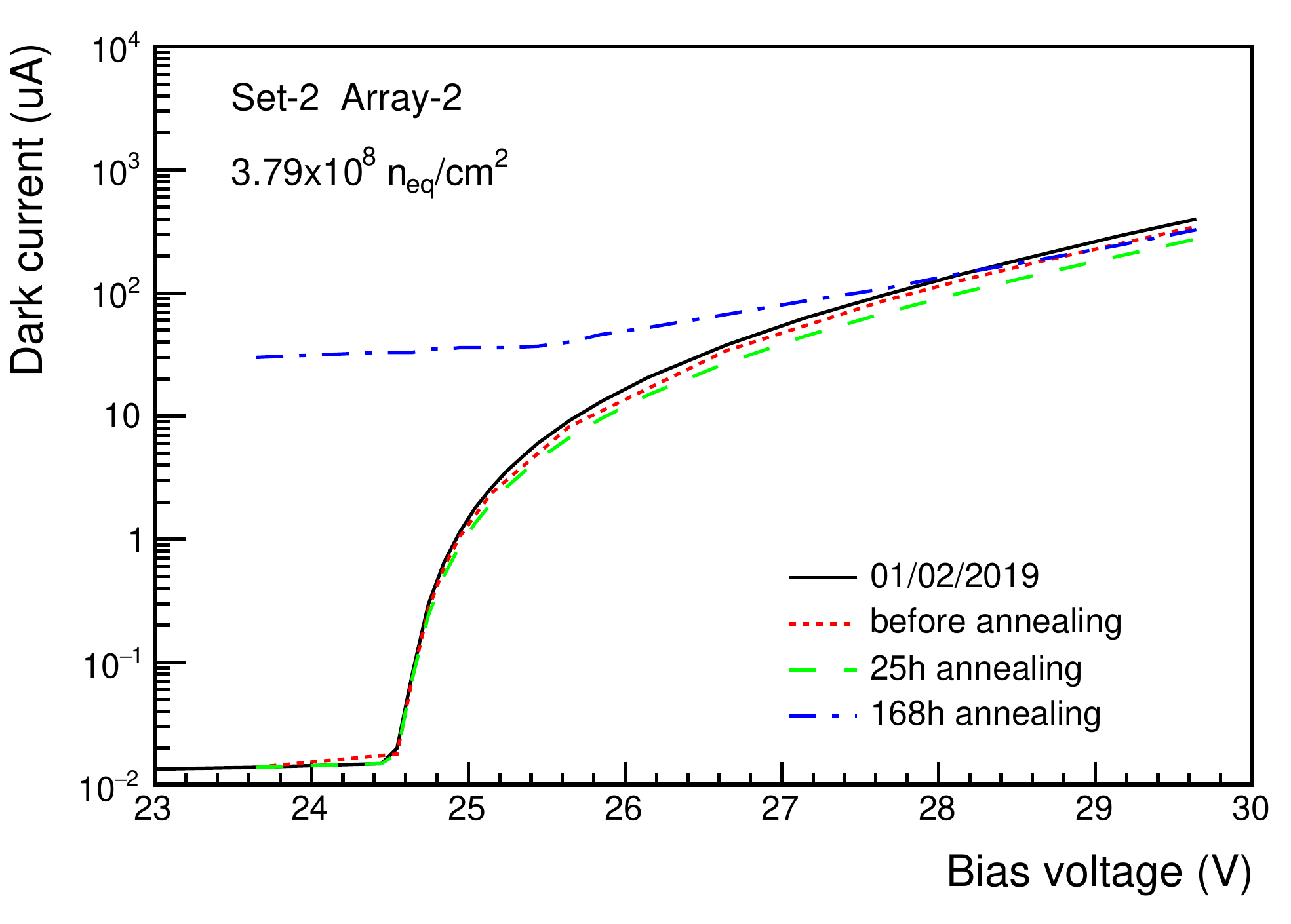}
    \includegraphics[width=0.45\textwidth]{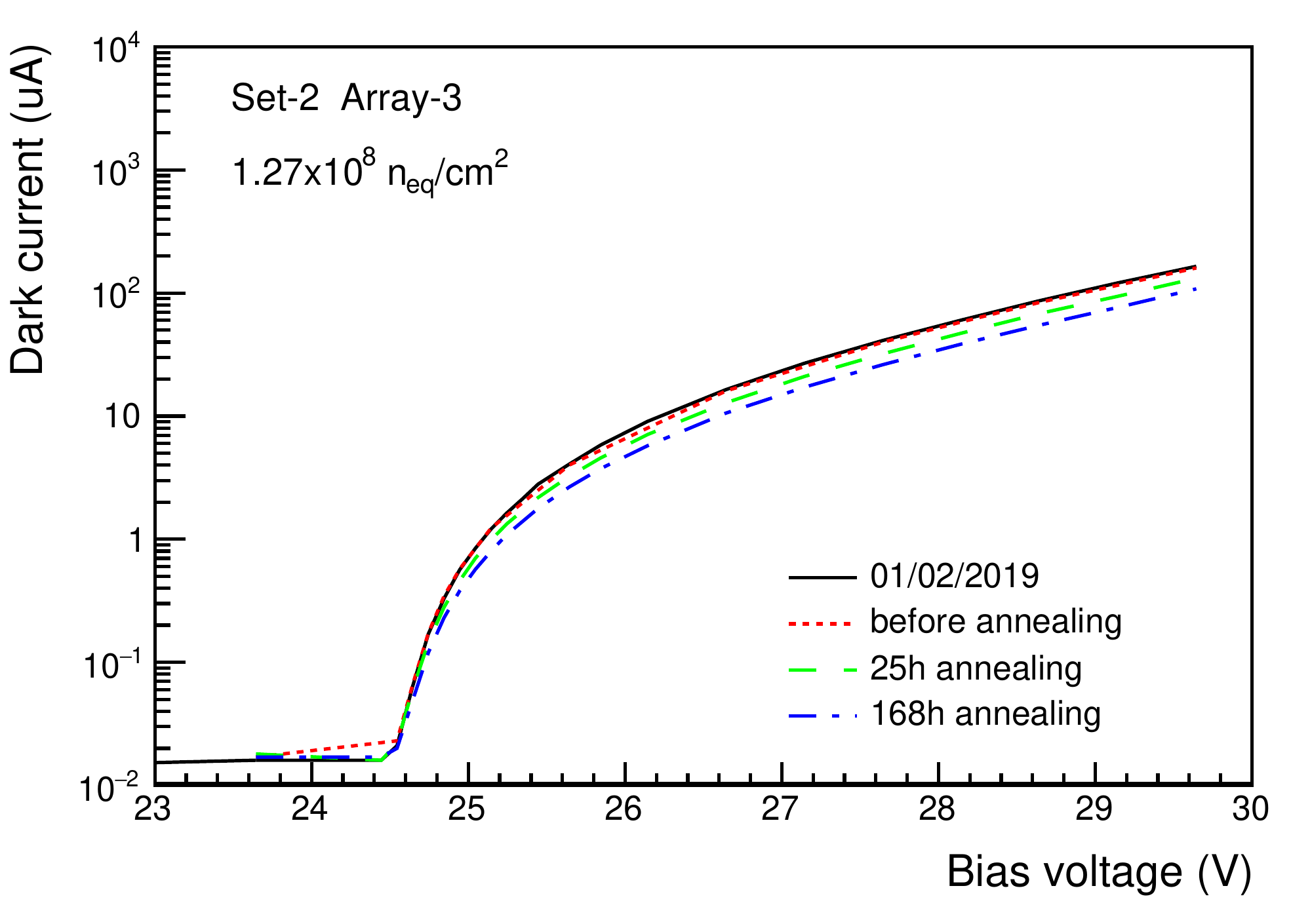}
    \includegraphics[width=0.45\textwidth]{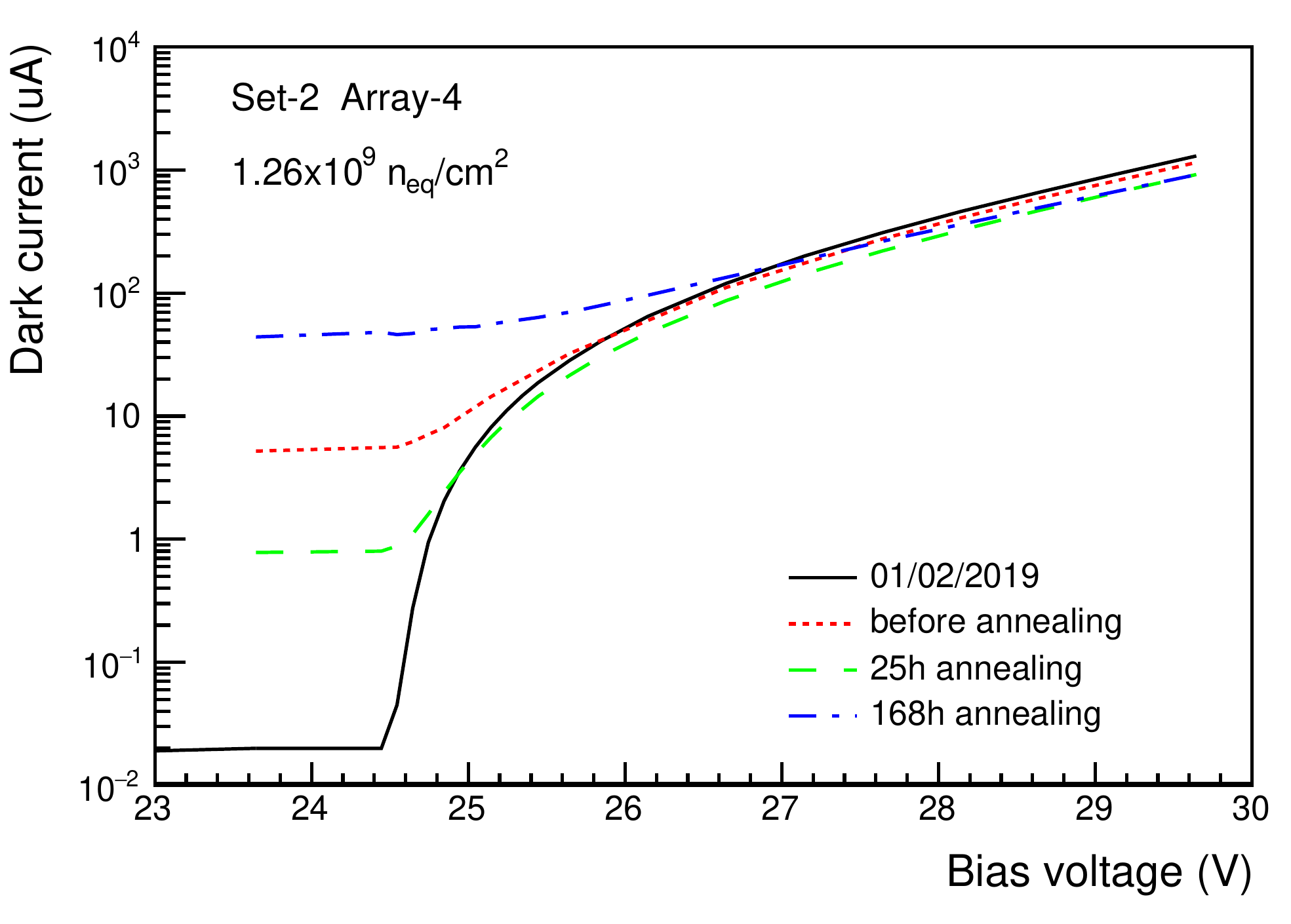}
    \caption{Current-voltage characteristics of several irradiated $2\times2$ SiPM arrays (1.47~cm$^2$) before and after annealing at 79$^\circ$C. }
    \label{fig:iv_annealing}
  \end{center}
\end{figure}

This annealing procedure yielded somewhat mixed results. Current-voltage curves of several SiPMs measured before and after annealing are shown in Figure~\ref{fig:iv_annealing}. All SiPMs demonstrated additional recovery of the displacement damage after each annealing stage, resulting in a lower amplified current (measured at a large overvoltage).
However, three out of the seven SiPM arrays from both sets showed a large increase in the non-amplified current (measured below the breakdown voltage). In some cases this non-amplified current significantly contributed to the total current above the breakdown voltage. This could be a result of mechanical damage sustained by these SiPMs from heating and/or insufficiently careful handling. Array-4 from \set{2} was already mechanically damaged before the annealing procedure (see Section~\ref{sec:iv}) and this damage was probably made worse by heating the sensor.

\begin{figure}[h]
  \begin{center}
    \includegraphics[width=0.8\textwidth]{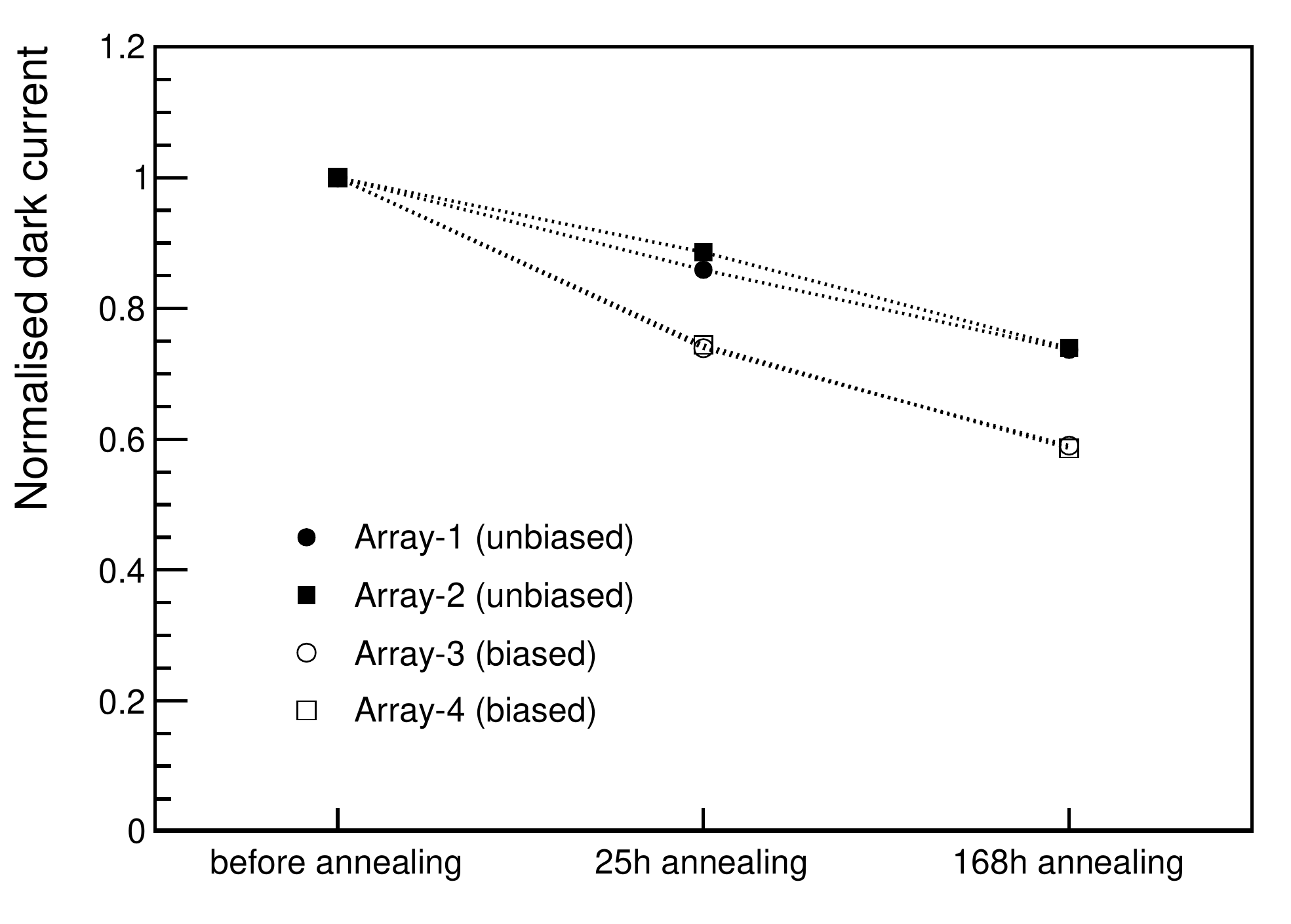}
    \caption{Dark current of $2\times2$ SiPM arrays from \set{1} at a bias voltage of 28.15~V measured before and after annealing at 79$^\circ$C. }
    \label{fig:annealing_current1}
  \end{center}
\end{figure}

\begin{figure}
  \begin{center}
    \includegraphics[width=0.8\textwidth]{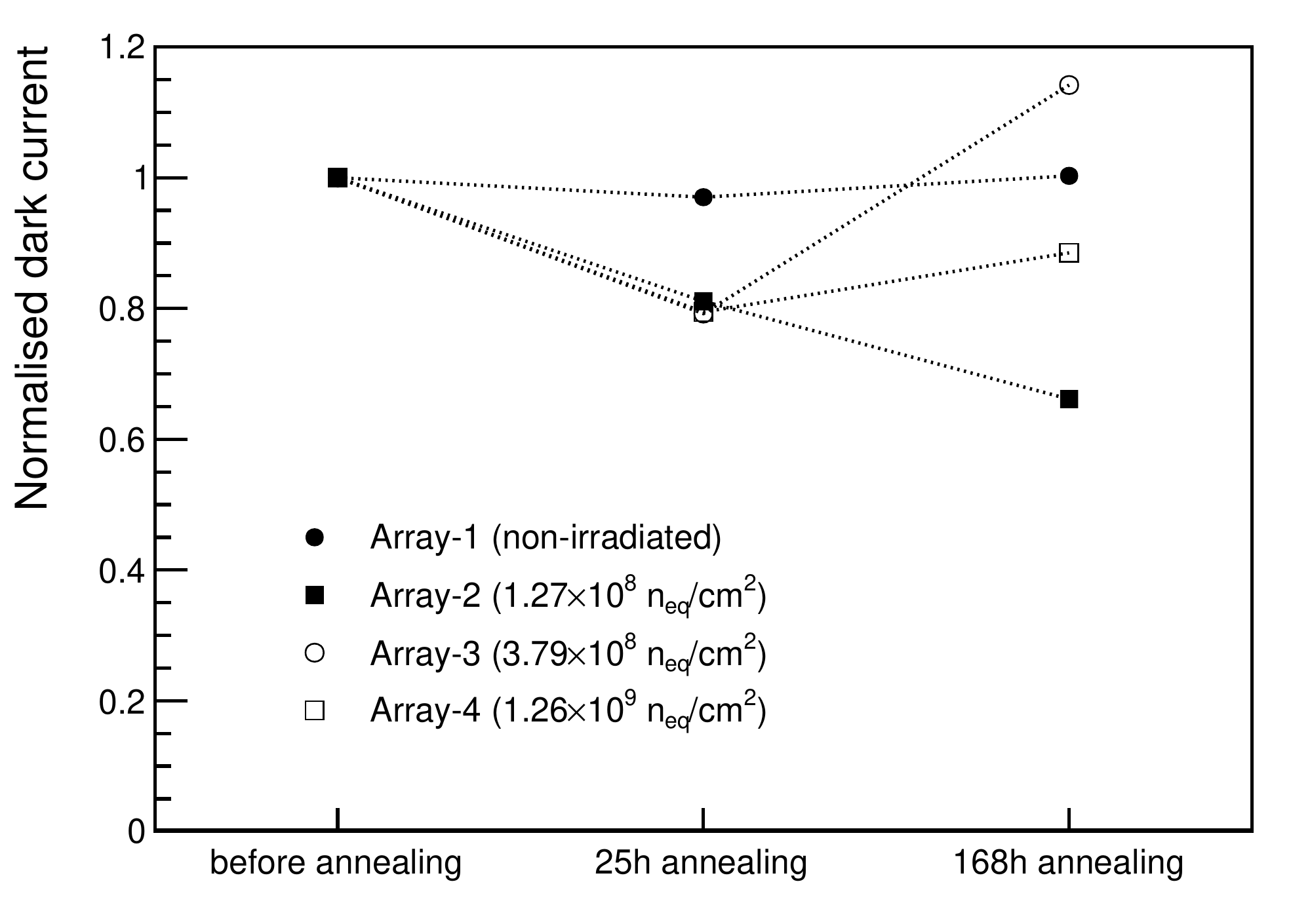}
    \caption{Dark current of $2\times2$  SiPM arrays from \set{2}  at a bias voltage of 28.15~V measured before and after annealing at 79$^\circ$C. }
    \label{fig:annealing_current2}
  \end{center}
\end{figure}

The relative change of the SiPM current at 28.15~V after the first and second annealing stage is shown in Figure~\ref{fig:annealing_current1} for the \set{1} SiPMs. The SiPM dark current decreased by about 14\% after the 25-hour unbiased annealing at 79$^\circ$C. A larger decrease of about 26\% was observed for the SiPMs annealed at a bias voltage of 24~V. After the total annealing period of 168 hours, the dark current decreased by 25\% for the unbiased SiPMs and by 41\% for the biased SiPMs.  Annealing of the \set{2} SiPMs yielded mixed results, reflecting damage suffered by SiPM-3 and SiPM-4 before and/or during the annealing procedure (Figure~\ref{fig:annealing_current2}).

The noise of \set{1} Array-3 and Array-4 annealed under bias voltage reduced by 14\% after the 24-hour annealing and by 24\% after 168-hour annealing. The  noise of \set{1} Array-1 and Array-2 annealed without bias reduced by 6\% after the 24-hour annealing and by 14\% after 168-hour annealing. The irradiated \set{2} SiPMs (also annealed without bias) showed a larger noise reduction of 11\% after the 24-hour annealing and by 17\% after 168-hour annealing.

\subsection{Temperature dependence of the SiPM dark current and noise}
\label{sec:temperature}

\begin{figure}
  \begin{center}
    \includegraphics[width=0.8\textwidth]{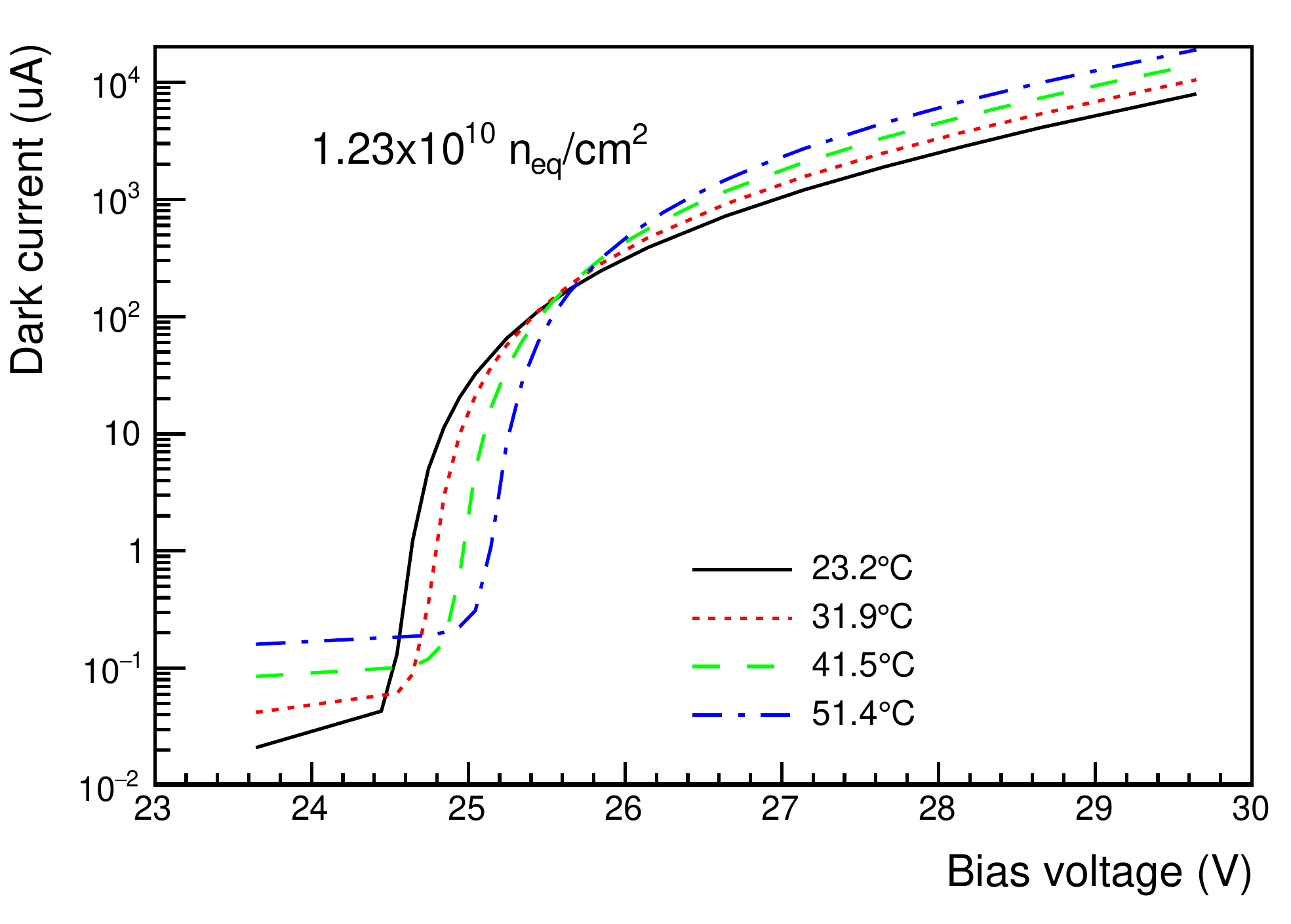}
    \caption{Current-voltage characteristics of a $2\times2$ SiPM array (Array-3 from \set{1}, 1.47~cm$^2$) measured at different temperatures.}
    \label{fig:temp_iv3}
  \end{center}
\end{figure}

Measurements of the SiPM dark current and noise at several different temperatures (23.2$^\circ$C, 31.9$^\circ$C, 41.5$^\circ$C and 51.4$^\circ$C) were performed on 16 May 2019 (that is after high temperature annealing of the SiPMs). These measurements were conducted for three  $2\times2$ SiPM arrays (non-irradiated, irradiated to $1.27\times10^{8}$ and $1.23\times10^{10}$~\ncm{}). In all three SiPM arrays, the breakdown voltage changed by 0.6$\pm$0.1~V when the temperature was increased by 28.2$^\circ$C as can be seen from the current-voltage curves in Figure~\ref{fig:temp_iv3}. This change is consistent with the temperature coefficient of 21.5~mV/$^\circ$C specified in the SiPM datasheet~\cite{jseries_ds}.

\begin{figure}
  \begin{center}
    \includegraphics[width=0.8\textwidth]{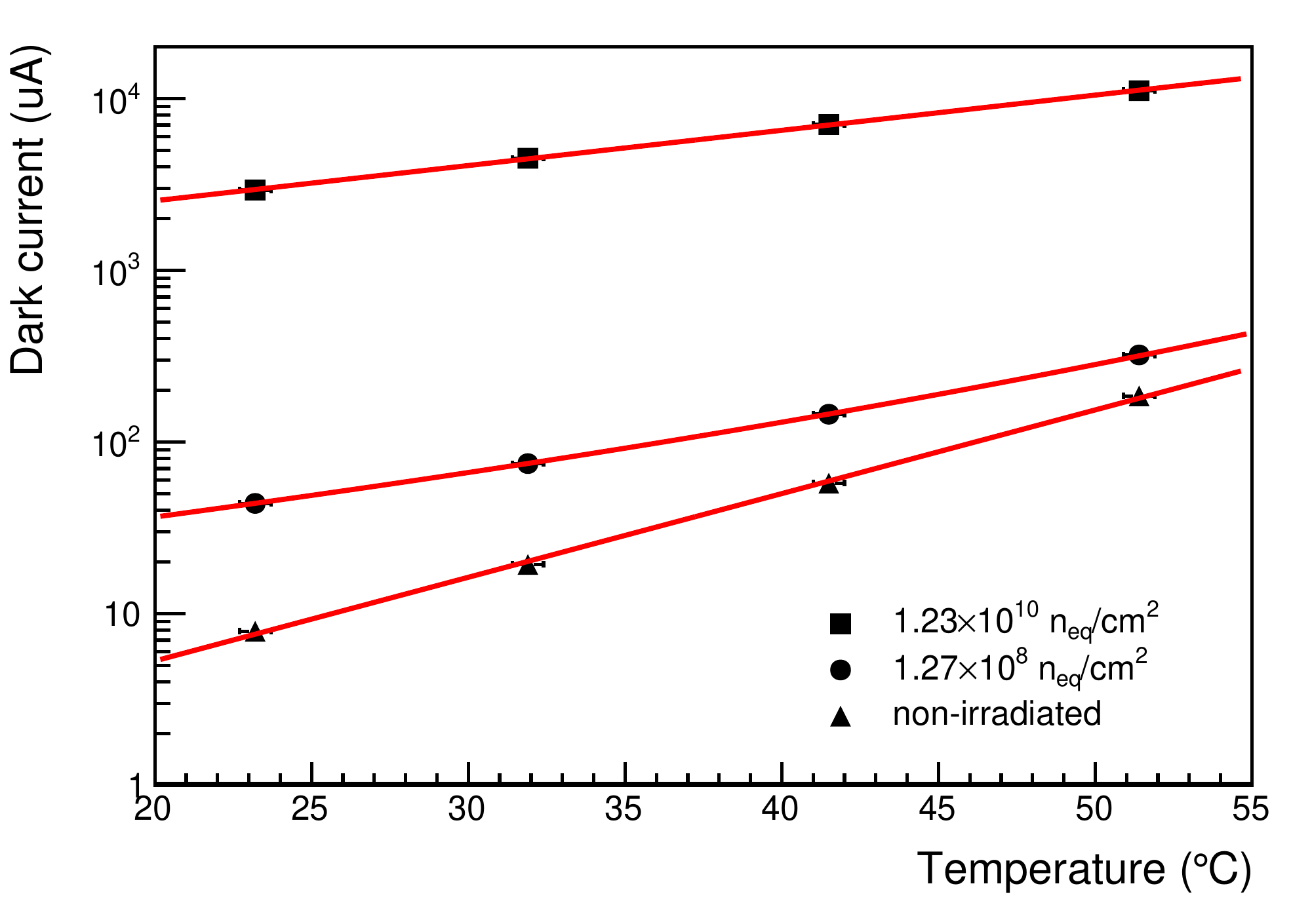}
    \caption{Dark current of $2\times2$ SiPM arrays  (1.47~cm$^2$) measured at different temperatures and a constant overvoltage of 3.65~V.}
    \label{fig:temp_current_cor}
  \end{center}
\end{figure}

Figure~\ref{fig:temp_current_cor} shows the SiPM dark current measured at different temperatures and a constant overvoltage of 3.65~V. Here, to maintain the constant overvoltage, the SiPM bias voltage was increased at a rate of 21.5~mV/$^\circ$C. For the non-irradiated $2\times2$ SiPM array, the temperature dependence of the dark current can be described by an exponential function $I_0(T)=\exp(-0.58+0.112T)$, where the current is given in uA and the temperature is in degrees $^\circ$C. For the SiPM array irradiated to $1.23\times 10^{10}$~\ncm{} (where the dark current is mainly due to the radiation damage), the temperature dependence is much weaker and is given by  $I_1(T)=\exp(6.89+0.047T)$. For the SiPM array irradiated to an intermediate fluence level of  $1.27\times 10^8$~\ncm{}, the current is described by a function $I(T) = I_0(T) + 0.0122 I_1(T)$, where $I_0$ is the current originally present in the SiPM array before the irradiation and the second term is a contribution from the radiation damage. The current $I_1$ is scaled down by a factor of 0.0122 to account for the lower irradiation fluence. This factor was obtained from fitting the function to the data and is somewhat larger than the factor of 0.0106 given by the ratio of the fluence levels.

\begin{figure}
  \begin{center}
    \includegraphics[width=0.8\textwidth]{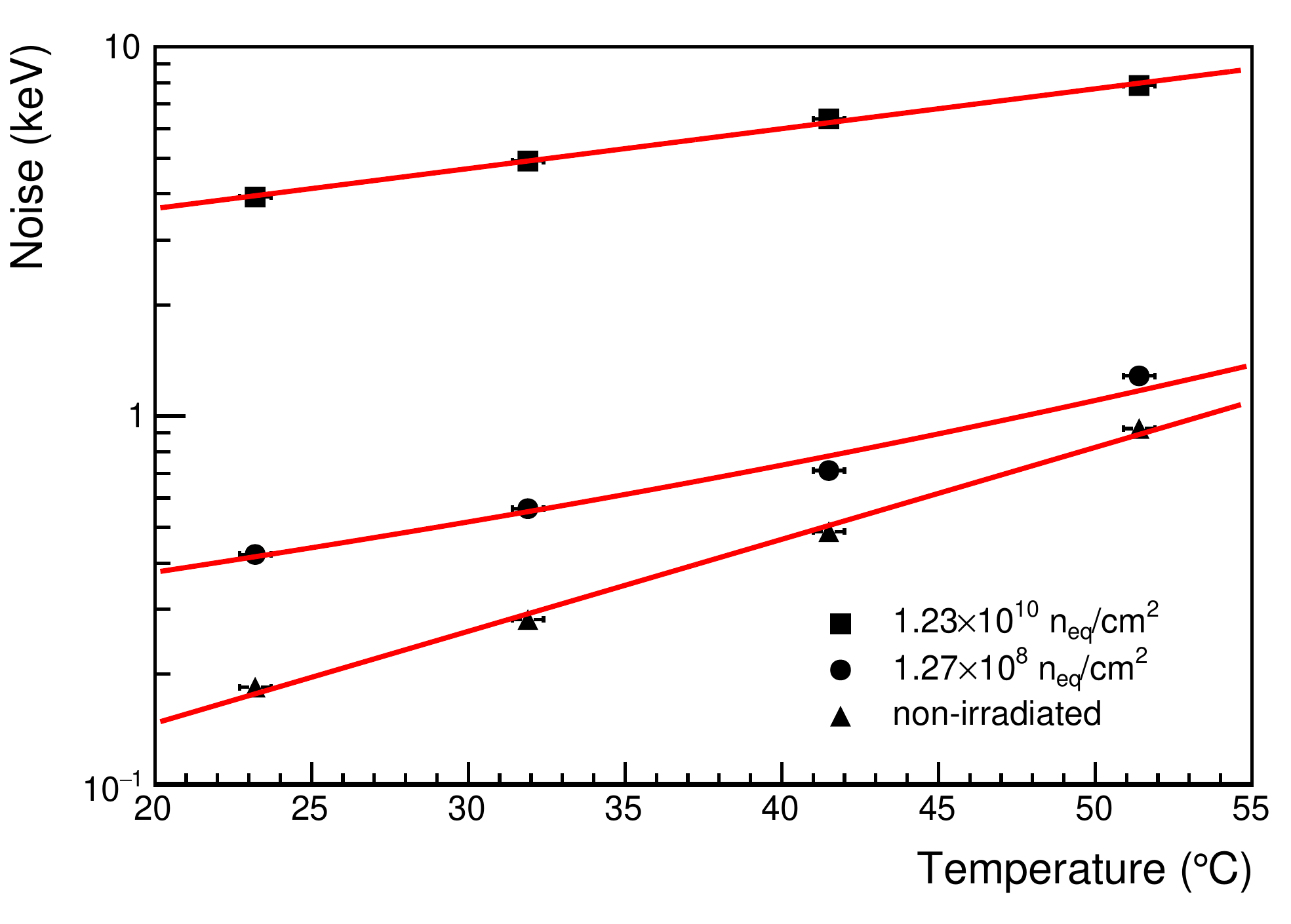}
    \caption{Dark noise of $2\times2$ SiPM arrays (1.47~cm$^2$) measured at different temperatures and a constant overvoltage of 3.65~V.}
    \label{fig:temp_noise_cor}
  \end{center}
\end{figure}

For a constant overvoltage of 3.65~V, the dark noise also grows exponentially with temperature: $\sigma_0(T) = \exp(-3.06+0.057T)$~keV for the non-irradiated SiPM array and $\sigma_1(T) = \exp(0.80+0.025T)$~keV for the one irradiated to $1.23\times 10^{10}$~\ncm{}  fluence (Figure~\ref{fig:temp_noise_cor}).

\section{Discussion}
\label{sec:discussion}

A large increase of the dark current observed in SensL J-series SiPMs after proton irradiation with 1~MeV neutron equivalent fluence ranging from  $1.27\times10^8$~\ncm{} to $1.23\times10^{10}$~\ncm{} is similar to the results obtained in other studies with different SiPM types~\cite{angelone2010,qiang2013,musienko2015,heering2016,li2016,lacombe2018}, although it is difficult to do a quantitative comparison because of different measurement conditions, such as time after irradiation, temperature and SiPM gain.

The reduction of the SiPM dark current by a factor of three observed 86 days after the irradiation suggests significant recovery from the proton induced radiation damage at room temperature. A similar self-annealing effect has been reported for early generations of SensL\footnote{A SPMArray unit based on ceramic design with 35 $\mu$m pixels -- Wafer Batch
Code X4151-05 using SPM3035 design.} and Hamamatsu\footnote{A preproduction unit of S10943-0258(X) MPPC with 50 $\mu$m pixels, equivalent to S12045.} SiPMs irradiated with neutrons to $3.7\times10^9$~\ncm~\cite{qiang2013}.
The dark current of the irradiated SiPMs left at room temperature decreased over time with a decay time constant of about 9 days. Over a period of 50 days it decreased by a factor of two.
Additional measurements showed that the recovery time strongly depends on temperature: annealing at $T=60^\circ$C reduced the recovery time to about 7 hours and allowed a larger reduction of the SiPM dark current. This trend was also observed in this study, where 25-hour annealing at 79$^\circ$C performed 154 days after the SiPM irradiation resulted in additional recovery of the SiPM dark current and noise. However, in contrast to the results of the previous study, further (although smaller) recovery was observed after 6 more days of annealing at 79$^\circ$C. It also should be noted that SiPMs biasing during the high temperature annealing resulted in larger recovery.

For all SiPM types, the dark current strongly depends on temperature. In this study, the dark current of the non-irradiated SiPM exponentially grew with increasing temperature at a rate of 11\%/$^\circ$C (at a constant overvoltage of 3.65~V). The dark current caused by radiation damage had a weaker temperature dependence of 4.7\%/$^\circ$C. This is similar to the results obtained for Hamamatsu SiPMs in Ref.~\cite{qiang2013}: a dark current increase of 9\%/$^\circ$C for non-irradiated SiPMs and an increase of 6.7\%/$^\circ$C after neutron irradiation.

Radiation damage of SiPMs in space is accumulated over a long time and the recovery process occurs simultaneously. Therefore, the results of the measurements performed in UCD after a three-month recovery period should better represent the SiPM damage expected in space than the results of the measurements in PSI. This is not quite true if SiPMs operate at much lower temperatures ($\ll 20^\circ$C) which may impede the recovery process~\cite{qiang2013}. The SiPM temperature on board EIRSAT-1 is expected to be in a range of $20-40^\circ$C. This means that the SiPM current and noise measured in UCD at $T=20^\circ$C may need to be scaled up by a factor of 2.5 and 1.6, respectively.

After irradiation to $1.23\times10^{10}$~n$_\mathrm{eq}$/cm$^2$ (corresponding to 6 years in a 550~km/40$^\circ$ orbit for the EIRSAT-1/GMOD configuration) and a recovery period of 86 days, the dark current of one 4-SiPM array (1.47~cm$^2$) at 21$^\circ$C was about 5~mA, or 3.4~mA/cm$^2$. This corresponds to a power consumption of approximately 100~mW per cm$^2$ of SiPM area, which may be prohibitive for a space instrument with a large number of SiPMs.
Moreover, such a large current may cause significant heating of SiPMs as experienced in this study, and a higher temperature would further increase the SiPM dark current and noise.

Operating SiPMs at a lower overvoltage may be considered to reduce the dark current. The recommended overvoltage range for J-series SiPMs is from 1 to 6~V~\cite{jseries_ds}. Operating SiPMs at an overvoltage of 1~V can reduce the dark current by a factor of 20 (compared to 3.65~V) at a cost of decreasing the gain by a factor of 3.65, some reduction in the photon detection efficiency (PDE) and increasing the SiPM sensitivity to variations of the breakdown voltage and operating temperature. Alternatively, SiPMs with a smaller microcell size can be used. The SiPM gain is proportional to the microcell capacitance and hence to the microcell area, therefore these SiPMs would generate a lower dark current after a similar radiation exposure. However, due to a smaller fill factor, such SiPMs typically have a substantially lower PDE, and hence a lower signal-to-noise ratio. Finally, where possible, SiPM cooling can significantly reduce both the dark current and noise, e.g. lowering the SiPM temperature by 30$^\circ$C would reduce the dark current of irradiated SiPMs by a factor of four and the noise by a factor of two.

For one 4-SiPM array (1.47~cm$^2$) in a CeBr3 gamma-ray detector maintained at 20$^\circ$C, the dark noise $\sigma$ grows from 0.2~keV before the irradiation to about 4~keV after the irradiation to an equivalent fluence of $1.23\times10^{10}$~n$_\mathrm{eq}$/cm$^2$ (Table~\ref{table:noise} in Section~\ref{sec:noise}). With an optimum SiPM readout, a threshold of about $4-5\sigma$ is needed to distinguish real gamma-ray events from false events produced by the SiPM noise. This means the gamma-ray detection threshold increases from about 1~keV to 20~keV. If a larger SiPM array is used, the noise and the detection threshold should be scaled proportionally to the square root of the number of SiPMs. It is clear that detection of low-energy gamma rays ($<$100~keV) may become problematic for long running high-altitude missions that want to use large SiPM arrays with monolithic scintillators, particularly for low light yield or very slow scintillators.

The above numbers are in a reasonable agreement with the gamma-ray spectral measurements conducted with the 4-SiPM and 16-SiPM CeBr3 detectors. It should be noted, however, that these measurements were affected by signal processing in the SIPHRA ASIC:
\begin{itemize}
 \item For the used CMIS attenuation of 100, the internal SIPHRA noise was comparable to the noise of the non-irradiated SiPMs, thereby somewhat increasing the detection threshold.
 \item The internal noise and the SiPM dark noise in the SIPHRA trigger channel (which is used to discriminate real events from noise based on the SiPM peak current) is not exactly the same as the noise measured in the spectroscopic channel.
 \item In the PSI measurements, the trigger for the 16-SiPM array was based on the excess current in any of the four sub-arrays, which is not as good as triggering on the total signal.
\end{itemize}

This study confirms that proton irradiation to a level of $1.23\times10^{10}$~n$_\mathrm{eq}$/cm$^2$ has no large effect on the PDE or gain of the J-series SiPMs. Although small changes of several percent cannot be excluded, the observed changes in detector response are probably due to temperature variations and certain effects of the readout ASIC.

SiPM radiation damage is not expected to be a problem for the EIRSAT-1 mission, as the SiPM exposure to radiation over the lifetime of one year will be limited to $4.3\times10^{8}$~n$_\mathrm{eq}$/cm$^2$. Depending on the SiPM temperature, the dark current of the entire 16-SiPM array may grow to 500-1200~uA. This current easily fits the mission power budget and will not have a large effect on the detector temperature. Although the detector noise will significantly increase throughout the mission lifetime, it will have little impact on the detection of gamma rays with energy above 50~keV that will be used for detection of gamma-ray bursts.

The only point of concern for the EIRSAT-1 gamma-ray detector will be a change in the scale and linearity of the detector response, as pulse-height measurements with SIPHRA are affected by the SiPM dark current and trigger threshold. Preliminary measurements confirm that the dark current compensation mode in SIPHRA (Section~\ref{sec:siphra}) greatly reduces the sensitivity of the trigger to average input current.
However, the relative positions of gamma-ray lines in measured spectra are still affected by proton radiation.   In space, the detector will be periodically calibrated using 511~keV emission from positron annihilation in Earth's atmosphere  but it may be impossible to maintain an accurate calibration at other energies. It should be noted, that a small calibration bias will have no significant effect on the instrument's ability to detect gamma-ray bursts, nor it will be a big factor in an analysis of statistically limited power-law GRB spectra.

\section{Conclusions}

SensL J-series MicroFJ-60035-TSV SiPMs were irradiated with 101.4~MeV protons with 1~MeV neutron equivalent fluence ranging from $1.27\times10^{8}$~n$_\mathrm{eq}$/cm$^2$ to $1.23\times10^{10}$~n$_\mathrm{eq}$/cm$^2$. Performance of these SiPMs was characterised in the context of using them as a readout for a CeBr3 gamma-ray detector in space applications. For SiPMs on board the EIRSAT-1 satellite, the exposure of $1.23\times10^{10}$~\ncm{} corresponds to 30 years in the ISS orbit or six years in a circular orbit at 550~km altitude and 40$^\circ$ inclination. Main conclusions from this study are summarised below.

Similar to other SiPMs, the J-series sensors experience a large increase in the dark current and noise after exposure to proton radiation at levels typical for space missions in low Earth orbits. Partial recovery of the radiation damage has been observed at room temperature. The increase of the dark current after irradiation and following room-temperature recovery has been found to be approximately proportional to the radiation fluence.

Assuming the same level of damage recovery occurs in space, the dark current of a single 0.37~cm$^2$ SiPM at room temperature and an overvoltage of 3.65~V increases to approximately 1.25~mA (3.4~mA/cm$^2$) after an exposure of $1.23\times10^{10}$~\ncm{}. This may be an issue for a long running high-altitude mission in terms of power consumption and thermal control.

After an exposure of $1.23\times10^{10}$~\ncm{}, the dark noise of a single 0.37~cm$^2$ SiPM in a CeBr3 detector at room temperature grows to $\sigma = 2$~keV. This increased noise pushes the gamma-ray detection threshold to $\sim$20~keV for a 4-SiPM (1.47~cm$^2$) CeBr3 detector and to $\sim$40~keV for a 16-SiPM (5.9~cm$^2$) detector at room temperature. The detector noise (expressed in keV) and gamma-ray detection threshold are inversely proportional to the photoelectron yield of the detector per unit energy and hence will be higher for less bright scintillators or where the emission spectrum does not match the peak spectral sensitivity of the SiPMs.

To decrease the dark current, operating SiPMs at a reduced overvoltage may be preferred. In addition, SiPM models operating with a lower gain, e.g. those having a small microcell size, may be considered. It should be noted, however, that SiPMs with a small microcell size typically have a lower photon detection efficiency (PDE) which can further degrade the signal-to-noise ratio of the detector and increase the gamma-ray detection threshold.

Where possible, the dark current and noise of SiPMs can be reduced by operating them at a lower temperature.
However, the dependence of the dark current and noise on temperature for irradiated SiPMs has been found to be substantially weaker than for non-irradiated SiPMs. Moreover, continuous operation at a low temperature may impede the recovery of SiPM radiation damage.

SiPM radiation damage will not be a problem for the EIRSAT-1 mission as the expected lifetime exposure is relatively low (about $4.3\times10^{8}$~\ncm{}) and the detector is only required to detect gamma-rays with energy above 50 keV.

\section*{Acknowledgements}
The test campaign was organised by the ESA Education Office in collaboration with the ESA Radiation Hardness Assurance and Component Analysis section (TEC-QEC) in the framework of the FYS programme. The authors are particularly grateful to Cesar Boatella Polo (TEC-QEC) and Alessandra Costantino (TEC-QEC) for their help in organising and conducting the irradiation test.

This study was supported under ESA contract 4000104771/11/NL/CBi. S.McB., J.M. and A.U. acknowledge support from Science Foundation Ireland under grant number 17/CDA/4723. D.M. acknowledges support from Irish Research Council under grant GOIPG/2014/453. The authors acknowledge support from SensL and IDEAS who provided several SiPM samples and the GALAO/SIPHRA kit free of charge.

\section*{References}

\bibliography{esa}

\end{document}